\documentclass[traditabstract]{aa}

\usepackage{graphicx}
\usepackage{txfonts}
\usepackage{natbib}
\usepackage[colorlinks=true,citecolor=blue]{hyperref}
\usepackage{xspace}
\usepackage{array}
\usepackage[labelformat=simple]{caption,subcaption}
\usepackage{amsmath,amssymb}
\usepackage{bm}
\usepackage{gensymb}

\bibpunct{(}{)}{;}{a}{}{,}

\newcommand{\as}{\hbox{$^{\prime\prime}$}\xspace}
\newcommand{\lsd}{\hbox{$\lambda/D$}\xspace}
\newcommand{\cpup}{c/p\xspace}
\newcommand{\degre}{\degree\xspace}

\defcitealias{N'Diaye2013}{Paper I}
\defcitealias{N'Diaye2016}{Paper II}
\defcitealias{Vigan2019}{Paper III}

\begin{document}

\title{Calibration of quasi-static aberrations in exoplanet direct-imaging instruments with a Zernike phase-mask sensor}
\subtitle{IV. Temporal stability of non-common path aberrations in VLT/SPHERE}
\titlerunning{Zernike phase-mask sensor. IV.}

\author{
  A. Vigan\inst{\ref{lam}}
  \and
  K. Dohlen\inst{\ref{lam}}
  \and
  M. N'Diaye\inst{\ref{oca}}
  \and
  F. Cantalloube\inst{\ref{lam}}
  \and
  J. Girard\inst{\ref{stsci},\ref{eso}}
  \and
  J. Milli\inst{\ref{ipag},\ref{eso}}
  \and
  J.-F. Sauvage\inst{\ref{lam},\ref{onera}}
  \and
  Z. Wahhaj\inst{\ref{eso}}
  \and
  G. Zins\inst{\ref{eso}}
  \and
  J.-L. Beuzit\inst{\ref{lam}}
  \and
  A. Caillat\inst{\ref{lam}}
  \and
  A. Costille\inst{\ref{lam}}
  \and
  J. Le Merrer\inst{\ref{lam}}
  \and
  D. Mouillet\inst{\ref{ipag}}
  \and
  S. Tourenq\inst{\ref{lam}}
}

\institute{
  Aix Marseille Univ, CNRS, CNES, LAM, Marseille, France \label{lam}
  \\ \email{\href{mailto:arthur.vigan@lam.fr}{arthur.vigan@lam.fr}}
  \and
  Universit\'e C\^ote d'Azur, Observatoire de la C\^ote d'Azur, CNRS, Laboratoire Lagrange, France \label{oca}
  \and
  Space Telescope Science Institute, Baltimore, MD, 21218, USA \label{stsci}
  \and
  ONERA, The French Aerospace Lab, BP72, 29 avenue de la Division Leclerc, 92322 Ch\^{a}tillon Cedex, France \label{onera}
  \and
  European Southern Observatory, Alonso de Cordova 3107, Vitacura, Santiago, Chile \label{eso}
  \and
  Univ. Grenoble Alpes, CNRS, IPAG, F-38000 Grenoble, France \label{ipag}
}
\date{Received 11 November 2021; accepted 9 February 2022}

\abstract{
  Coronagraphic imaging of exoplanets and circumstellar environments using ground-based instruments on large telescopes is intrinsically limited by speckles induced by uncorrected aberrations. These aberrations originate from the imperfect correction of the atmosphere by an extreme adaptive optics (ExAO) system; from static optical defects; or from small opto-mechanical variations due to changes in temperature, pressure, or gravity vector. More than the speckles themselves, the performance of high-contrast imagers is ultimately limited by their temporal stability, since most post-processing techniques rely on difference of images acquired at different points in time. Identifying the origin of the aberrations and the timescales involved is therefore crucial to understanding the fundamental limits of dedicated high-contrast instruments. In previous works we demonstrated the use of a Zernike wavefront sensor called ZELDA for sensing non-common path aberrations (NCPA) in the VLT/SPHERE instrument. We now use ZELDA to investigate the stability of the instrumental aberrations using five long sequences of measurements obtained at high cadence on the internal calibration source. Our study reveals two regimes of decorrelation of the NCPA. The first, with a characteristic timescale of a few seconds and an amplitude of a few nanometers, is induced by a fast internal turbulence within the enclosure. The second is a slow quasi-linear decorrelation on the order of a few $10^{-3}$\,nm\,rms/s that acts on timescales from minutes to hours. We use coronagraphic image reconstruction to demonstrate that these two NCPA contributions have a measurable impact on differences of images, and that the fast internal turbulence is a dominating term over to the slow linear decorrelation. We also use dedicated sequences where the derotator and atmospheric dispersion compensators emulate a real observation to demonstrate the importance of performing observations symmetric around the meridian, which minimizes speckle decorrelation, and therefore maximizes the sensitivity to point sources in difference of images.
}


\maketitle

\section{Introduction}
\label{sec:introduction}

High-contrast imagers on large ground-based telescopes, such as SPHERE on the VLT \citep{Beuzit2019}, GPI on Gemini South \citep{Macintosh2014}, and SCExAO on Subaru \citep{Jovanovic2015}, are designed to suppress stellar light and reveal the faint signal of nearby planetary companions \citep[e.g.,][]{Macintosh2015,Chauvin2017,Keppler2018} or circumstellar disks \citep[e.g.,][]{Boccaletti2015,Kalas2015,Garufi2017}. These instruments combine extreme adaptive optics (ExAO) systems \citep[e.g.,][]{Guyon2005b,Fusco2006} with efficient coronagraphs \citep[e.g.,][]{Soummer2005,Guyon2005a} and advanced post-processing methods \citep{Racine1999,Marois2006,Lafreniere2007,Cantalloube2015,Ruffio2017}. They typically reach contrasts of $10^{-5}$--$10^{-6}$ at a 400-500\,mas \citep{Langlois2021}, or even a few $10^{-7}$ in the most favorable cases \citep{Vigan2015}, which is sufficient to detect young giant exoplanets and set constraints on their population through large-scale surveys \citep{Nielsen2019,Vigan2021}.

The ultimate contrast performance is limited by the small amplitude and phase errors in the optical system, which vary over time and create the quasi-static speckle field in the focal-plane coronagraphic images. In ground-based systems these errors come from the imperfect correction of the atmosphere by the ExAO system \citep{Macintosh2005,Hinkley2007}, which typically have variation timescales of a few seconds; from effects that are not detectable by the wavefront sensor (WFS), such as the low-wind effect \citep{Sauvage2015,Milli2018}; from effects that are introduced by the ExAO system itself, such as the (asymmetric) wind-driven halo \citep{Cantalloube2018,Cantalloube2020}; and finally from temporal variations inside the instrument itself (temperature, flexures, moving optics, Fresnel propagation), which create aberrations that are not all seen by the wavefront sensor of the ExAO system and are therefore not corrected \citep{Sauvage2007}.

The last are called the non-common path aberrations (NCPA) and are the primary source of speckles in the focal plane in good observing conditions. Many different strategies have been proposed to calibrate the NCPA in hardware or in software, and in either a static or dynamic way \citep[e.g.,][]{Galicher2010,N'Diaye2013,Paul2013,Martinache2013,Martinache2014,Wilby2017,Bottom2017,Skaf2021}. The variation timescales of the NCPA is therefore an important topic, which was identified very early on as the main limitation in high-contrast imaging \citep[e.g.,][]{Sivaramakrishnan2002,Soummer2007}. Most post-processing techniques rely on some degree of correlation between the science images being analyzed and some reference images. The reference images can be acquired either on the science target itself with angular diversity \citep[ADI;][]{Marois2006} or spectral diversity \citep[SDI;][]{Racine1999}, or using reference stars observed close in time \citep[e.g.,][]{Wahhaj2021} or for which data is available in large online archives \citep[e.g.,][]{Choquet2016}.

The decorrelation of NCPA has been the topic of many studies either in the laboratory \citep{Martinez2012,Martinez2013} or on sky \citep{Hinkley2007,Milli2016}. All of these studies share an important aspect: they were all conducted on focal-plane coronagraphic images and tried to quantify the decorrelation time of speckles based on focal-plane metrics. Moreover, the on-sky studies did not try to disentangle the different sources that create the quasi-static speckles, in relation with their respective timescales.

We propose here a different approach based on a Zernike wavefront sensor (\citealt{N'Diaye2013}, hereafter \citetalias{N'Diaye2013}), which measures small phase errors in a diffraction-limited high-contrast imaging system. A prototype of this sensor called the Zernike sensor for Extremely Low-level Differential Aberration (ZELDA) was installed in VLT/SPHERE \citep{N'Diaye2014,N'Diaye2016spie}, and we have already demonstrated its use for calibration of NCPA both on the internal source of the instrument (\citealt{N'Diaye2016}, hereafter \citetalias{N'Diaye2016}) and on sky (\citealt{Vigan2019}, hereafter \citetalias{Vigan2019}). Thanks to its sensitivity and the way it is implemented in SPHERE, ZELDA is able to measure aberrations with nanometric accuracy in the pupil plane, with a spatial resolution up to 192\,cycles/pupil (\cpup), which is significantly higher than what the ExAO system can correct (up to 20\,\cpup). This approach enables the quantitative determination of the NCPA variation over time, of the spatial frequencies that are the most impacted by the NCPA, and of which moving optical elements impact this variation the most.

In the present work we explore the most fundamental limit of the contrast in VLT/SPHERE. In Sect.~\ref{sec:temporal_variation} we study the temporal variations of the NCPA in a fully static configuration of the instrument and we look at their impact on the contrast reached in difference of focal-plane coronagraphic images. Then in Sect.~\ref{sec:impact_derotator} we study the impact of the derotator, which is one of the important contributors in NCPA variations during real observations, and in Sect.~\ref{sec:realistic_observing_sequence} we consider an even more realistic observing sequence that mimics a real observation. Finally, we conclude and discuss our results in Sect.~\ref{sec:conclusions_discussion}.

\section{Temporal variations of the NCPA}
\label{sec:temporal_variation}

\begin{figure*}
  \centering 
  \includegraphics[width=1\textwidth]{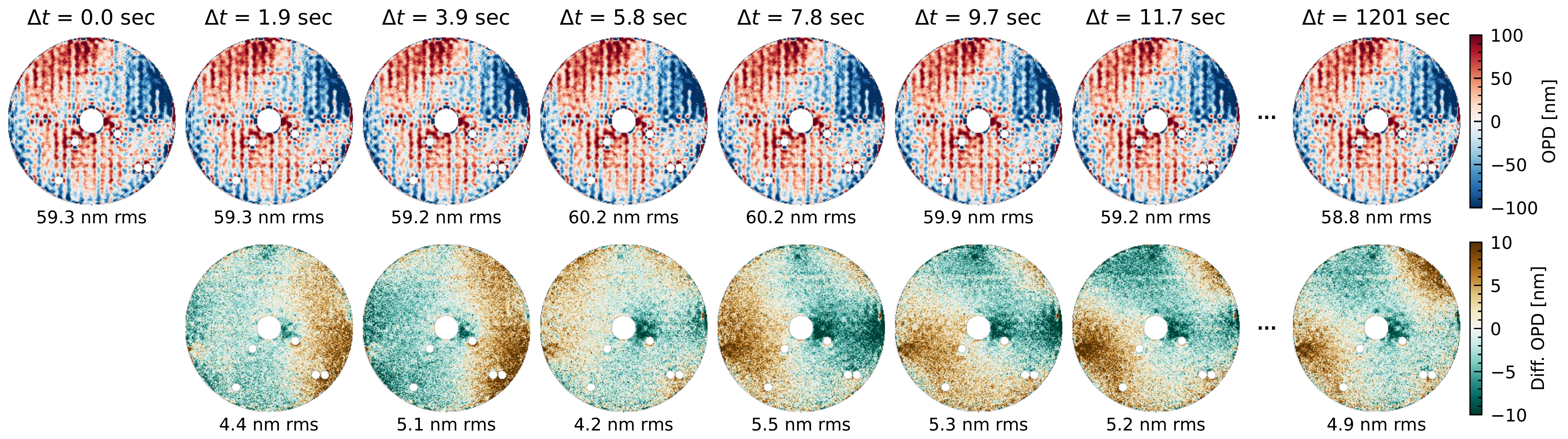}
  \caption{Temporal gallery of eight OPD maps acquired on 2015 December 16. The first seven maps are consecutive, while the last one was acquired after approximately 20\,min. The top row shows the individual OPD maps and the bottom row shows the differential maps obtained by subtracting the first map to the subsequent ones. The dead actuators of the deformable mirror and the region corresponding to the central obscuration have been masked. The standard deviation is reported underneath each map.}
  \label{fig:temporal_gallery}
\end{figure*}

The temporal evolution of the NCPA in SPHERE was previously estimated to be a few tenths of a nanometer over periods of 1\,h, based on the decorrelation of speckles in focal-plane coronagraphic images \citep{Martinez2013}. However, this estimate must be taken with caution because the translation between speckle decorrelation, or loss of contrast, and NCPA variations is far from direct and relies on several theoretical assumptions. Moreover, the data of \citet{Martinez2013} were acquired under laboratory conditions during the final performance assessment of SPHERE in Europe. After the instrument was installed at the telescope, another study focused on the decorrelation of speckles was performed by \citet{Milli2016} and revealed at least two timescales at play in the focal-plane coronagraphic data.

We now describe using ZELDA to measure the slow temporal variations of the NCPA. For this purpose, we acquired five long sequences of data at high cadence on the internal source, which enable measuring the evolution of the wavefront over timescales of a few hours. In this study we focus on the evolution up to 1.5\,hours, which corresponds to the typical duration of an ADI sequence used in recent large-scale surveys such as GPIES \citep{Nielsen2019} and SHINE \citep{Vigan2021}. In Sect.~\ref{sec:experimental_data_temporal} we present the details of the data acquisition, in Sect.~\ref{sec:temporal_decorrelation} we discuss the temporal evolution of the NCPA based on our measurements, and finally in Sect.~\ref{sec:impact_coronagraphic_images_temporal} we quantify the impact on reconstructed coronagraphic images.

\subsection{Experimental data}
\label{sec:experimental_data_temporal}

\begin{table}
  \caption[]{Temporal sequences on the internal source}
  \label{tab:ncpa_temporal_sequences}
  \centering
  \begin{tabular}{lcccc}
    \hline\hline
    Date          & DIT & $N_{img}$ & Duration & Precomp.\tablefoottext{a}{} \\
                  & [s] &          & [min]    & \\
    \hline                                    
    2015-12-16    & 1.0 & 3500     & 116      & No  \\
    2015-12-24    & 1.0 & 5000     & 166      & Yes \\
    2017-03-25\,a & 1.0 & 6000     & 198      & No  \\
    2017-03-25\,b & 1.0 & 6000     & 198      & No  \\
    2020-02-10    & 1.0 & 6000     & 198      & Yes \\
    \hline
  \end{tabular} 
  \tablefoot{\tablefoottext{a}{This column indicates whether the static NCPA have been pre-compensated using a ZELDA closed loop before the acquisition of the sequence (see \citetalias{N'Diaye2016} and \citetalias{Vigan2019})}.}
\end{table}

We recall that in SPHERE the ZELDA phase mask is located in the near-infrared (NIR) coronagraphic wheel. With the mask placed into the NIR beam, we used a pupil imaging lens in the IRDIS instrument \citep{Dohlen2008} to acquire the data with the corresponding Hawaii-2RG detector. We also recall that a ZELDA measurement requires three types of data to produce optical path difference (OPD) maps calibrated in nanometers: data obtained with the mask inserted into the beam, clear pupil data obtained without the mask, and finally instrumental background data. These data are then fed to the \texttt{pyZELDA}\footnote{\url{https://github.com/avigan/pyZELDA}} \citep{Vigan2018pyzelda} code to produce OPD maps. We refer the reader to \citetalias{N'Diaye2013} for the ZELDA formalism, and to \citetalias{N'Diaye2016} and \citetalias{Vigan2019} for the details on the data acquisition, analysis, and performance.

The five sequencies presented in this work were acquired on the internal source of the instrument over a period of a few years (Table~\ref{tab:ncpa_temporal_sequences}). We used a detector integration time (DIT) of 1\,s, which, combined with the detector overheads, provides a new image every $\Delta t = 1.938$\,s. A slightly higher cadence could have been obtained with the minimum DIT of 0.84\,s, but the amount of flux in the ZELDA images would have been decreased by 15\%, therefore decreasing the signal-to-noise ratio (S/N) of the measurements. To avoid any additional noise, 100~DIT of 1\,s were acquired for the clear pupil and instrumental background images. Two different sequences were acquired on 2017 March 25 (hereafter 2017-03-25\,a and 2017-03-25\,b) with an interruption of approximately 1\,hour between them, due to other activities with the telescope.

In some cases (see Table~\ref{tab:ncpa_temporal_sequences}) the NCPA of the instrument were pre-compensated using the closed-loop procedure described in \citetalias{N'Diaye2016} and \citetalias{Vigan2019}. This procedure was not applied systematically, but depended on the other tests that were performed just before the acquisition of the ZELDA sequence. This should not have an impact on the results presented here. The ZELDA sensor dynamic range slightly decreases as the amount of NCPA increases, but here we are in a regime of low NCPA (50--70\,nm\,rms, see \citetalias{Vigan2019}) where the sensitivity of the sensor is very high. As demonstrated in \citetalias{Vigan2019}, the reconstructed OPD maps underestimate the true aberrations in the instrument by $\sim$20\%, so a sensitivity loss factor equal to 0.8 must be taken into account. This corresponds to the optical gains of the sensor in the presence of aberrations \citep{Chambouleyron2020}: the sensitivity of the sensor is influenced by the amount of aberrations in the system, so this needs to be calibrated and taken into account with a gain factor. From here on, all the stated results take into account the correction factor.

\subsection{Temporal decorrelation}
\label{sec:temporal_decorrelation}

\subsubsection{Pair-wise analysis}
\label{sec:pairwise_analysis}

In Figure~\ref{fig:temporal_gallery} we show in the top row a sequence of eight OPD maps obtained on 2015 December 16, and in the bottom row the residuals after subtracting the first OPD map from the subsequent ones. The top row reveals that the main structures of the NCPA, such as the imprint of the DM actuators and the large-scale instrumental aberrations, remain quasi-static over timescales of at least 20 minutes. However, the differential OPD maps in the bottom row show a different picture with aberrations on the order of a few nanometers (root mean square, rms) varying over timescales of a few seconds.

\begin{figure}
  \centering 
  \includegraphics[width=0.49\textwidth]{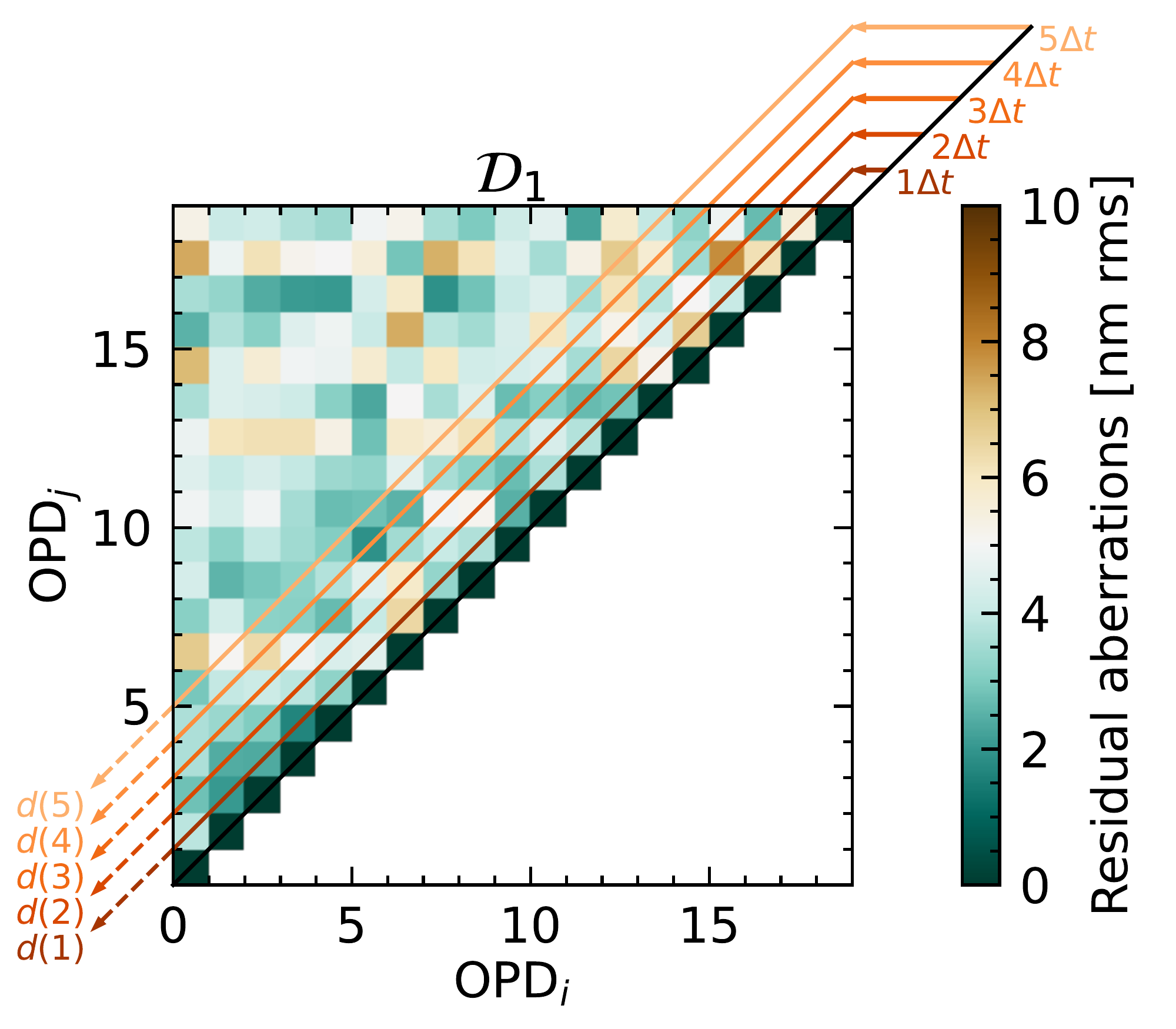}
  \caption{Illustration of a small 20$\times$20 decorrelation matrix $\mathcal{D}_1$ for 20 consecutive OPD maps extracted from the 2017-02-23\,b sequence. The value of each point $(i,j)$ of the matrix corresponds to the amount of aberrations between 1 and 2\,\cpup in $\delta\mathrm{OPD}_{ij} = \mathrm{OPD}_i - \mathrm{OPD}_j$ (see Sect.~\ref{sec:pairwise_analysis} for details). Each diagonal ($1\Delta t$, $2\Delta t$, ...) corresponds to a fixed time difference between OPDs. The average decorrelation $d(t)$ of the OPD maps is computed by averaging values along each diagonal.}
  \label{fig:decorrelation_matrix}
\end{figure}

\begin{figure}
  \centering 
  \includegraphics[width=0.49\textwidth]{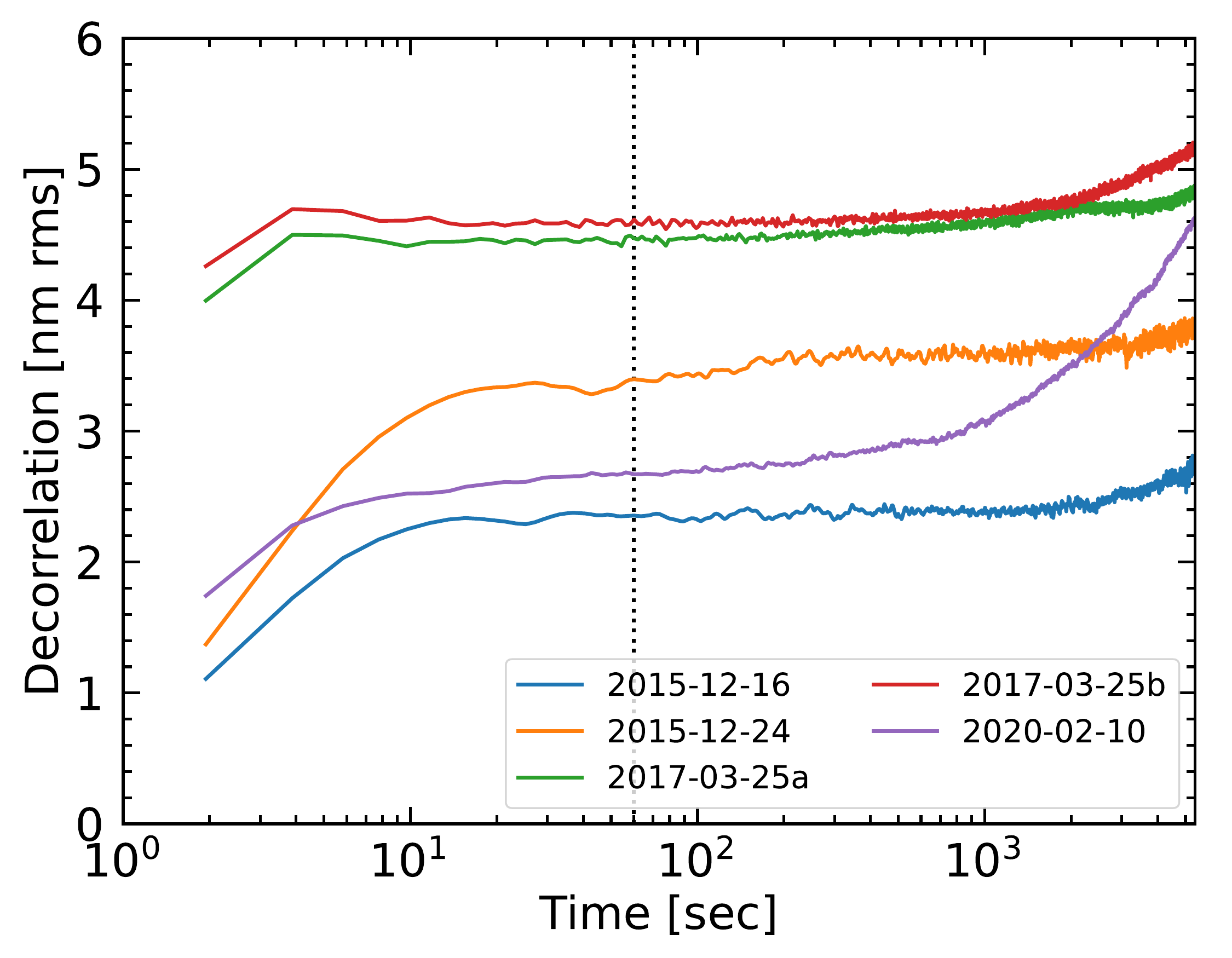}
  \caption{Average decorrelation of OPD maps between 1 and 2\,\cpup over time differences up to 1.5\,hours in our sequences. The curves are computed as the average along the diagonals of $\mathcal{D}_f$ (i.e., at fixed time differences, see Sect.~\ref{sec:temporal_decorrelation}). Two distinct regimes are visible: a fast decorrelation on timescales below a few seconds and a much slower, quasi-linear decorrelation on timescales of minutes to hours. The dotted line at 60\,s corresponds to the limit for the fitting of the parameters of the decorrelation (Eq.~\ref{eq:decorr}).}
  \label{fig:temporal_decorrelation_example}
\end{figure}

For a more quantitative estimate of the variations of the NCPA and the timescales involved, we performed pair-wise subtraction between all of the OPD maps acquired in the sequences listed in Table~\ref{tab:ncpa_temporal_sequences}. For each sequence of $N$ OPD maps, we computed the differential OPD maps $\delta\mathrm{OPD}_{ij} = \mathrm{OPD}_i - \mathrm{OPD}_j$ for $i = 1 \ldots N$ and $j = 1 \ldots i$. Then we computed the power spectral density (PSD) of $\delta\mathrm{OPD}_{ij}$ and used the standard deviation $\sigma(f)$ defined in \citetalias{Vigan2019} to calculate the amount of residual aberrations in discrete bins of spatial frequency from 1 to 50\,\cpup. We recall the definition of $\sigma(f)$,

\begin{equation}
  \sigma(f) = \sqrt{\int_{f}^{f+1}\int_{0}^{2\pi}\mathrm{PSD}(\nu, \theta)\nu d\nu d\eta},
\end{equation}

\noindent with $\nu$ and $\eta$ being the radial and azimuthal coordinates of the spatial frequencies in the 2D PSD. The values of $\sigma(f)$ are expressed in nm rms and represent the integrated quantity of aberrations in the bin of spatial frequencies between $f$ and $f+1$ cycles/pupil. These values are easier to relate to wavefront errors than the classical PSD expressed in (nm~/~(\cpup))$^2$.

At the end of the process, for each spatial frequency $f$, we obtain a decorrelation matrix $\mathcal{D}_f$ of dimensions $N \times N$ that represents the amount of differential aberrations between any two OPD maps in the sequence. A simple illustration based on only 20 OPD maps extracted from the 2017-03-25\,b sequence is provided in Fig.~\ref{fig:decorrelation_matrix}.

Since the time step $\Delta t$ between each OPD map is constant in a sequence, moving from the central diagonal towards one of the corners of $\mathcal{D}_f$ is equivalent to moving in constant time steps. We therefore have $N-1$ differences of OPD maps separated by $\Delta t$, $N-2$ separated by $2\Delta t$, $N-3$ separated by $3\Delta t$, and so on. This large sample of differential OPD maps enables us to look robustly at the average decorrelation of NCPA as a function of time by computing statistics along the diagonals of $\mathcal{D}_f$. The fixed time step diagonals are illustrated in Fig.~\ref{fig:decorrelation_matrix}.

\begin{figure*}
  \centering 
  \includegraphics[width=1\textwidth]{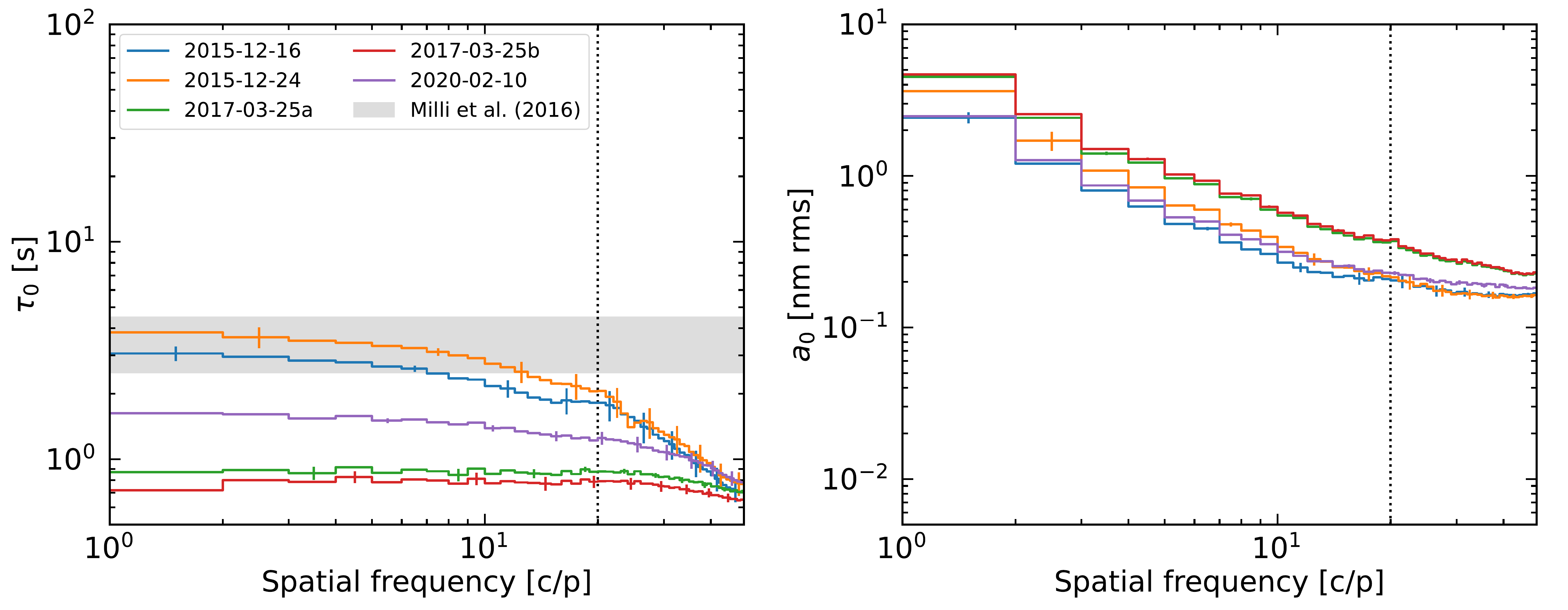}
  \caption{Fit of the parameters modeling the fast exponential decorrelation of the NCPA as a function of spatial frequency in the original sequences. The quantities $\tau_0$ and $a_0$ represent the characteristic time and amplitude of the decorrelation. The dotted line at 20\,\cpup shows the cutoff of the SPHERE ExAO system. The range of values estimated by \citet{Milli2016} from focal-plane images are overplotted in gray for comparison.}
  \label{fig:temporal_decorrelation_expfit_turb=1}
\end{figure*}

The average decorrelation as a function of time computed for all sequences is plotted in Fig.~\ref{fig:temporal_decorrelation_example}. Two regimes of decorrelation are clearly identified. The first is a fast decorrelation that occurs on timescales of a few seconds, and the second is much slower quasi-linear decorrelation that occurs on timescales of several minutes. These two regimes of decorrelation $d$ likely have distinct origins so we model them separately as a function of time $t$. We do not aim to perform a detailed physically based modeling of $d$, but simply to estimate its main properties in terms of timescale and amplitude. This is why we use empirical modeling:
\begin{align}
  \label{eq:decorr}
  d(t) =
  \begin{cases}
    a_0\left(1-e^{-\frac{t}{\tau_0}}\right) + s_0t & \text{for } t \leq 60\,s \\
    s_1t + a_1 & \text{for } t > 60\,s
  \end{cases}.
\end{align}
\noindent Here $\tau_0$ and $a_0$ are the characteristic time and amplitude of the fast decorrelation, respectively, and $s_1$ is the slope of the slow decorrelation that occurs over long timescales. We also include an additional linear slope term $s_0$ in the modeling of the fast initial decorrelation because we found that it improves the fit to the data. This is, however, an empirical finding that we do not investigate further. Similarly, we do not report the values for $a_1$ because the main parameter of interest is the decorrelation rate over long timescales, $s_1$.

From the physical point of view, a turbulent process with a characteristic time $\tau$ can be expected to average out over time with an exponential decrease, which is what we observe here with the fast decorrelation. On longer timescales, processes related to thermal variations and flexures, which are assumed to be the main cause of NCPA variations, are also expected to be random but on very long timescales. They could therefore be modeled with a decreasing exponention, but with a timescale so long that it would appear linear over the duration of our observations. The linear variation of NCPA with time is routinely used to model the evolution of aberrations for high-contrast space missions \citep[e.g.,][]{Pogorelyuk2020}.

For the actual estimation of the parameters we use a Levenberg-Marquardt minimization algorithm from the \texttt{scipy.optimize} package \citep{SciPy2020}. The fit is performed on our five sequences and independently on each spatial frequency bin up to 50\,\cpup. We estimate the error bars by performing the fit for $\tau_0$ and $a_0$ over the first 10 to 60\,s in steps of 5\,s, and for $s_1$ with an upper limit on $t$ from 1000 to 5400\,s in steps of 50\,s. The error bars reflect the standard deviation of values obtained for all the fits.

\subsubsection{Fast internal turbulence}
\label{sec:fast_internal_turbulence}

The results of the fast decorrelation fit using Eq.~\ref{eq:decorr} are presented in Fig.~\ref{fig:temporal_decorrelation_expfit_turb=1}. The initial decorrelation occurring on timescales of a few seconds can be interpreted as a fast internal turbulence within the enclosure of the instrument. \citet{Milli2016} had previously reported a fast decorrelation of the speckles in focal-plane images acquired with SPHERE/IRDIS, both on sky and on the internal source, but they were unable to explain the origin of this decorrelation. With the present data we can confidently identify internal turbulence in the non-common path as the origin of the speckle decorrelation that they observed.

The origin of this fast internal turbulence was investigated soon after its discovery in 2016. The first suspect was the air flow that is constantly pumped into the enclosure of the instrument to maintain a clean environment. However, several tests performed with and without the air flow did not show any change in the amplitude of the internal turbulence. The second suspect was the motor of the apodizer wheel. This wheel is located in a pupil plane between the two sets of prisms of the NIR atmospheric dispersion corrector (ADC). The motor is placed exactly underneath the optical beam, downstream of the visible-NIR dichroic filter, which means that any aberrations instroduced at that location are not seen by the ExAO WFS. Measurements performed with a thermal camera during the integrations of SPHERE in Europe in 2014 show that its temperature is several degrees above the rest of the SPHERE bench. It is therefore extremely likely that the motor is heating the air just at the level of the NIR beam and creates the observed turbulence. A possible mitigation would be the installation of a copper braid between the motor and the main bench structure to evacuate some of the heat. However, the motor is inconveniently placed and the overall amount of aberrations due to the fast internal turbulence was not deemed sufficient to require a specific intervention at the time.

The characteristic timescale for the fast internal turbulence, $\tau_0$, varies from $\sim$0.7 (or less) to $\sim$3.5\,s over the five sequences. For the two 2017 March 25 sequences the measured timescale (0.7--0.9\,s) is shorter than the time resolution of the data (1.9\,s), so this estimation is an upper limit. However, the tail of the exponential is visible at short timescales in Fig.~\ref{fig:temporal_decorrelation_example} and allows us to perform a fit, although the accuracy may be worse than for the other sequences. Generally, there seems to be a decrease in $\tau_0$ with the spatial frequency, but the magnitude of this decrease varies from one sequence to another. For example the two 2017 March 25 sequences show almost no dependence of $\tau_0$ on the spatial frequency, while the 2015 December 24 sequence shows a decrease of a factor 3 between 1\,\cpup and 30\,\cpup. This is qualitatively similar to the variations with angular separation observed by \citet{Milli2016}. The values that they observed (4.5--2.5\,s) are high compared to most of our sequences, but since our analysis shows a strong variability between epochs it remains compatible with two of our sequences at spatial frequencies below 10\,\cpup.

The amplitude of the decorrelation, $a_0$, is similar for our different sequences, with a spread of a factor $<$2. This observation reinforces the hypothesis of a common physical origin from the apodizer wheel motor for this turbulence. The amplitude is higher at low spatial frequencies and follows a linear decrease in log-log space. The amplitude of the decorrelation induced by the fast internal turbulence is on the order of 5\,nm\,rms, which remains small with respect to the amplitude of the total NCPA ($\sim$60\,nm\,rms). It is, however, the same order of magnitude as the value expected for the slow linear decorrelation of the aberrations that occurs over longer timescales, as we show in the next sections.

\subsubsection{Subtraction of the fast internal turbulence}
\label{sec:subtraction_internal_turbulence}

\begin{figure}
  \centering 
  \includegraphics[width=0.49\textwidth]{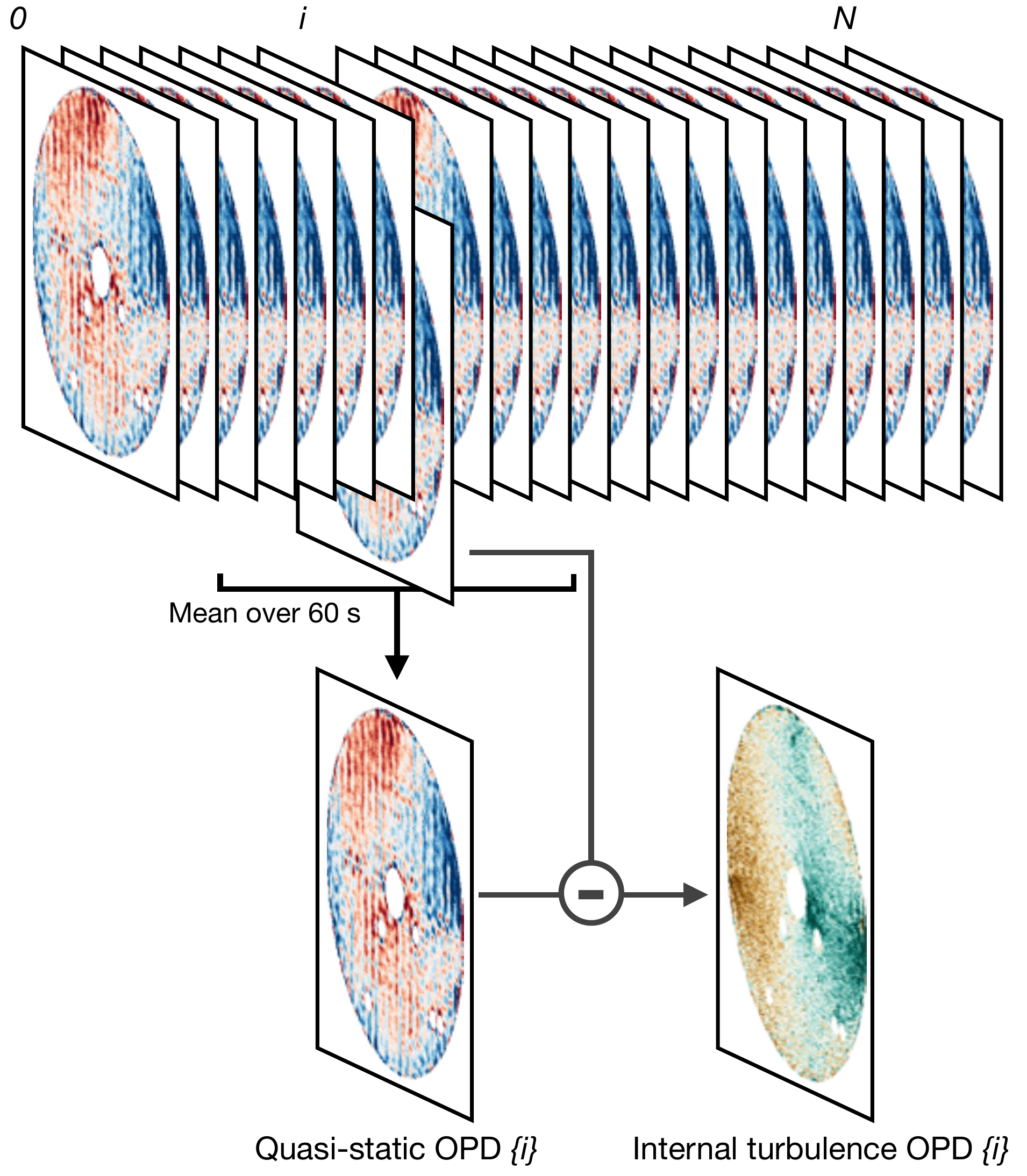}
  \caption{Illustration of how the quasi-static and fast internal turbulence components of the NCPA are estimated for each image $i$ of an OPD map sequence.}
  \label{fig:internal_turbulence_subtraction}
\end{figure}

\begin{figure*}
  \centering 
  \includegraphics[width=1\textwidth]{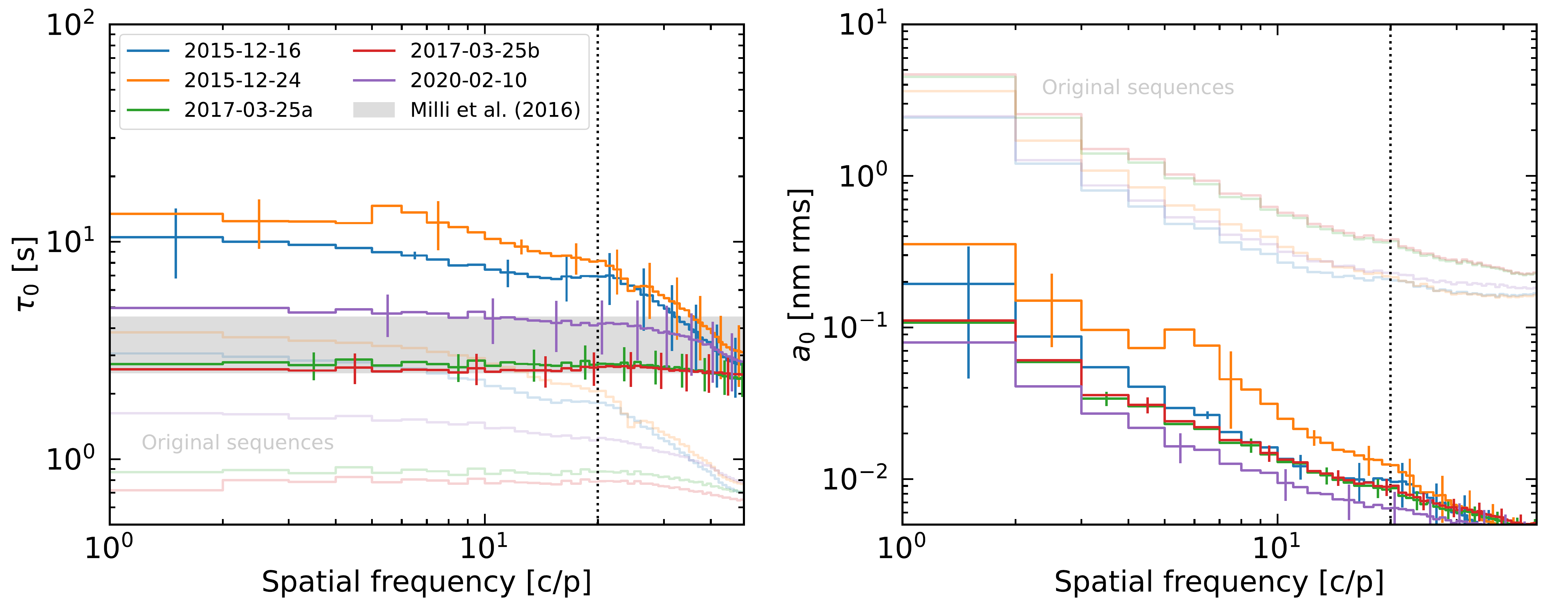}
  \caption{Fit of the parameters modeling the fast exponential decorrelation of the NCPA as a function of spatial frequency in the turbulence-subtracted sequences (see Sect.~\ref{sec:subtraction_internal_turbulence} for details). The faint overplotted curves correspond to the original sequences presented in Fig.~\ref{fig:temporal_decorrelation_expfit_turb=1} to demonstrate the effect of the internal turbulence subtraction.}
  \label{fig:temporal_decorrelation_expfit_turb=0}
\end{figure*}

From Fig.~\ref{fig:temporal_decorrelation_example} we qualitatively infer that the amplitude of the fast decorrelation and the long-term slow decorrelation can reach similar levels. Although this figure only shows the results in the 1--2\,\cpup spatial frequency bin, the same trend is observed at higher spatial frequencies. This means that our estimation of the slow decorrelation of the NCPA over long timescales may be biased by the fast internal turbulence.

To quantify this bias, we estimate the static and quasi-static parts of the NCPA using a sliding mean of the OPD maps over a duration of 60\,s\footnote{The results presented hereafter do not strongly depend on the choice of this value, as long as it is more than a few dozen seconds.}, which is equivalent to a temporal low-pass filtering of the data. With a characteristic timescale $\tau_0$ of typically a few seconds or less, we assume that the fast internal turbulence will mostly average out over 60\,s. Then we isolate the OPD induced by the fast internal turbulence by subtracting the sliding mean in each individual OPD map. The process is illustrated in Fig.~\ref{fig:internal_turbulence_subtraction}. The different NCPA terms and their impact on coronagraphic images is studied in the following sections.

To check whether the subtraction of the fast internal turbulence was successful, we performed again the pair-wise analysis described in Sect.~\ref{sec:pairwise_analysis} on the turbulence-subtracted sequences. The new results for the fast decorrelation are presented in Fig.~\ref{fig:temporal_decorrelation_expfit_turb=0}. After subtracting the turbulence from the data, the characteristic time $\tau_0$ was increased from 0.7--3.5\,s to 3--13\,s in the 1--5\,\cpup range, which corresponds to a factor $\sim$4 extension. The fact that $\tau_0$ did not increase by an even higher factor probably implies that the turbulence was not fully subtracted. However, and more importantly, the amplitude of the decorrelation uniformly decreased by factor 12 to 47 depending on the sequence. This is a massive decrease in amplitude that confirms that the effect of the fast internal turbulence was significantly attenuated in the data. We therefore used these new sequences to study the slow, long-term evolution of the NCPA described in the next section.

For the remainder of the paper we refer to the sequences that still contain the fast internal turbulence as the original sequences, and the sequences where it has been subtracted as the turbulence-subtracted sequences.

\subsubsection{Slow linear decorrelation}
\label{sec:slow_linear_decorrelation}

\begin{figure}
  \centering
  \includegraphics[width=0.49\textwidth]{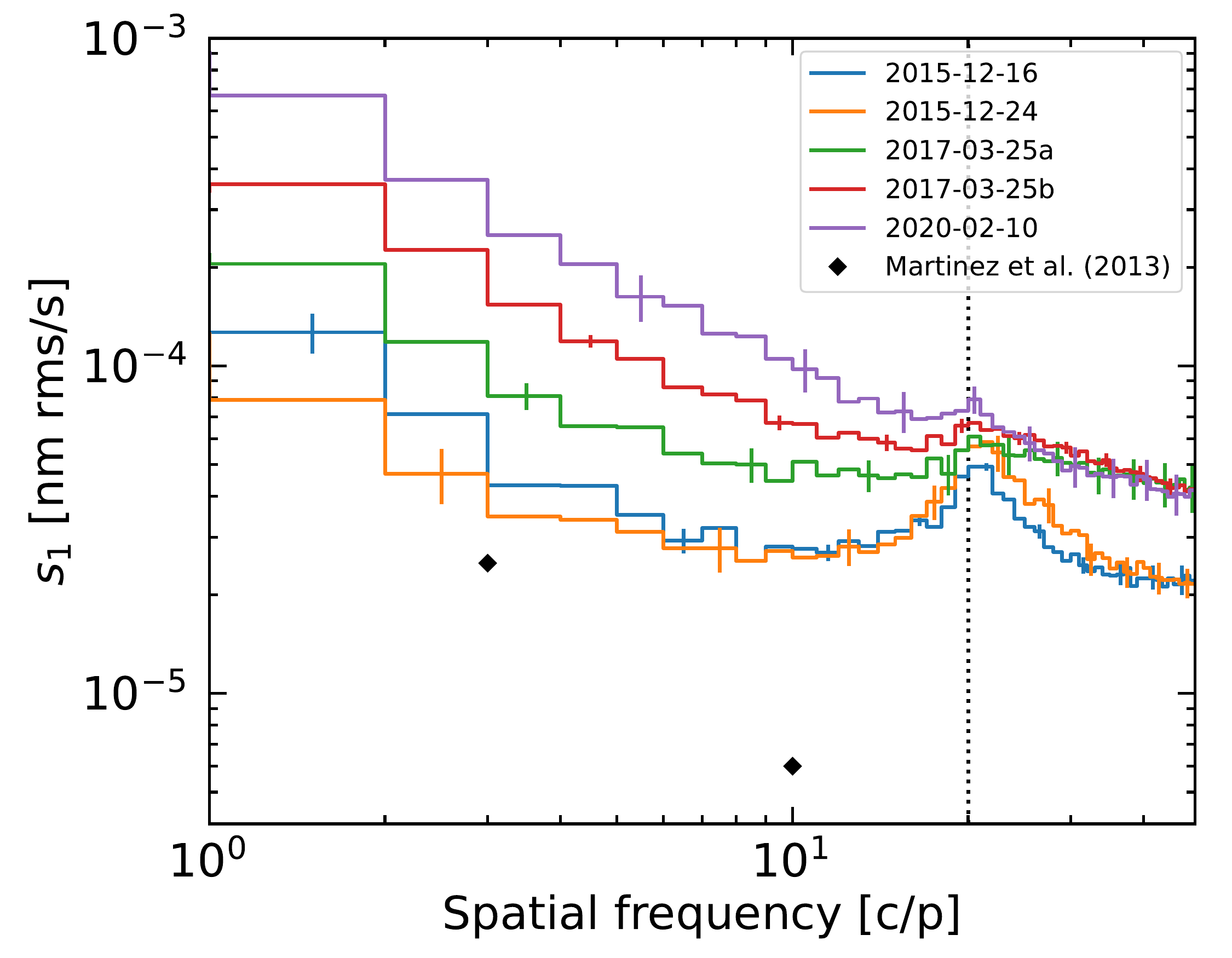}
  \caption{Fit of the parameter $s_1$ modeling the long-term decorrelation rate of the NCPA as a function of spatial frequency. The dotted line at 20\,\cpup shows the cutoff of the SPHERE ExAO system. The values inferred by \citet{Martinez2013} from speckle decorrelation in focal-plane coronagraphic images are overplotted with black diamond symbols.}
  \label{fig:temporal_decorrelation_linfit_turb=0}
\end{figure}

The rate of the slow decorrelation that occurs on timescales longer than a few minutes, $s_1$, varies significantly between our five sequences (Fig.~\ref{fig:temporal_decorrelation_example}). This is confirmed in Fig.~\ref{fig:temporal_decorrelation_linfit_turb=0}, which shows the estimated value of $s_1$ for the different sequences as a function of spatial frequency. Over the five other sequences, the value of the decorrelation rate $s_1$ differs by up to almost one order of magnitude at low spatial frequencies and by a factor lower than 2 when reaching 20\,\cpup. Between 1 and 10\,\cpup, the integrated value of the decorrelation ranges from $3.60 \times 10^{-5}$ to $3.14 \times 10^{-4}$\,nm\,rms/s, which linearly translates from 0.19 to 1.70\,nm\,rms on a timescale of 1.5\,hours. Again these values are small with respect to the total amount of NCPA, but not negligible.

There appears to be a trend with time, with older sequences (2015) showing a lower rate of decorrelation than the more recent ones (2020). The 2020 February 10 sequence already appears to be a particularly extreme case in Fig.~\ref{fig:temporal_decorrelation_example}, with a decorrelation rate at least twice as high as any other sequence. Additional sequences would be needed to understand if we are witnessing some long-term degradation or if the observed variability is caused by other factors. A peculiar effect is clearly identified for all sequences at the 20\,\cpup cutoff frequency of the AO system, which indicates that the ExAO system has an effect on the observed slow decorrelation. One hypothesis could be that the amplitude of the decorrelation depends on the time elapsed between the daily AO calibrations and the acquisition of the ZELDA sequences. This will require further investigation in the future.

We can compare these estimations to the work of \citet{Martinez2013}, who inferred a value for the NCPA decorrelation based on the decorrelation of the speckles in focal-plane images when the SPHERE instrument was being tested in Europe in 2013. The value that they measured at 3\,\cpup is slightly lower than our sequence showing the lowest decorrelation rate (2015-12-24) at that spatial frequency but remains compatible given the large variability that we observe. However, their value at 10\,\cpup is lower by a factor of $\sim$4.5 with respect to the 2015 December 24 sequence. Their values are approximately a factor 5 below the average of our own estimations, but the slope of the decrease between their measurements at 3 and 10\,\cpup is similar to the value we observe. This could denote a common origin in the decorrelation of these systematic, which are generally associated with the variations in the environmental conditions (temperature, pressure, flexures), but with a higher initial level of systematics. Their values are to be taken with caution because their approach is much less straightforward and involves important assumptions to link a contrast loss with a variation in the physical level of aberrations. They are not able to verify that their theoretical prescriptions are valid based on real data, which we consider as a limitation in their work. Of course we cannot guarantee that our own analysis is perfect, but going from wavefront errors to coronagraphic images is much more straightforward than the opposite. Moreover, our previous results with ZELDA have already demonstrated a high level of accuracy in reproducing SPHERE data, both on the internal source or on sky \citepalias{N'Diaye2016,Vigan2019}.

\begin{figure}
  \centering
  \includegraphics[width=0.49\textwidth]{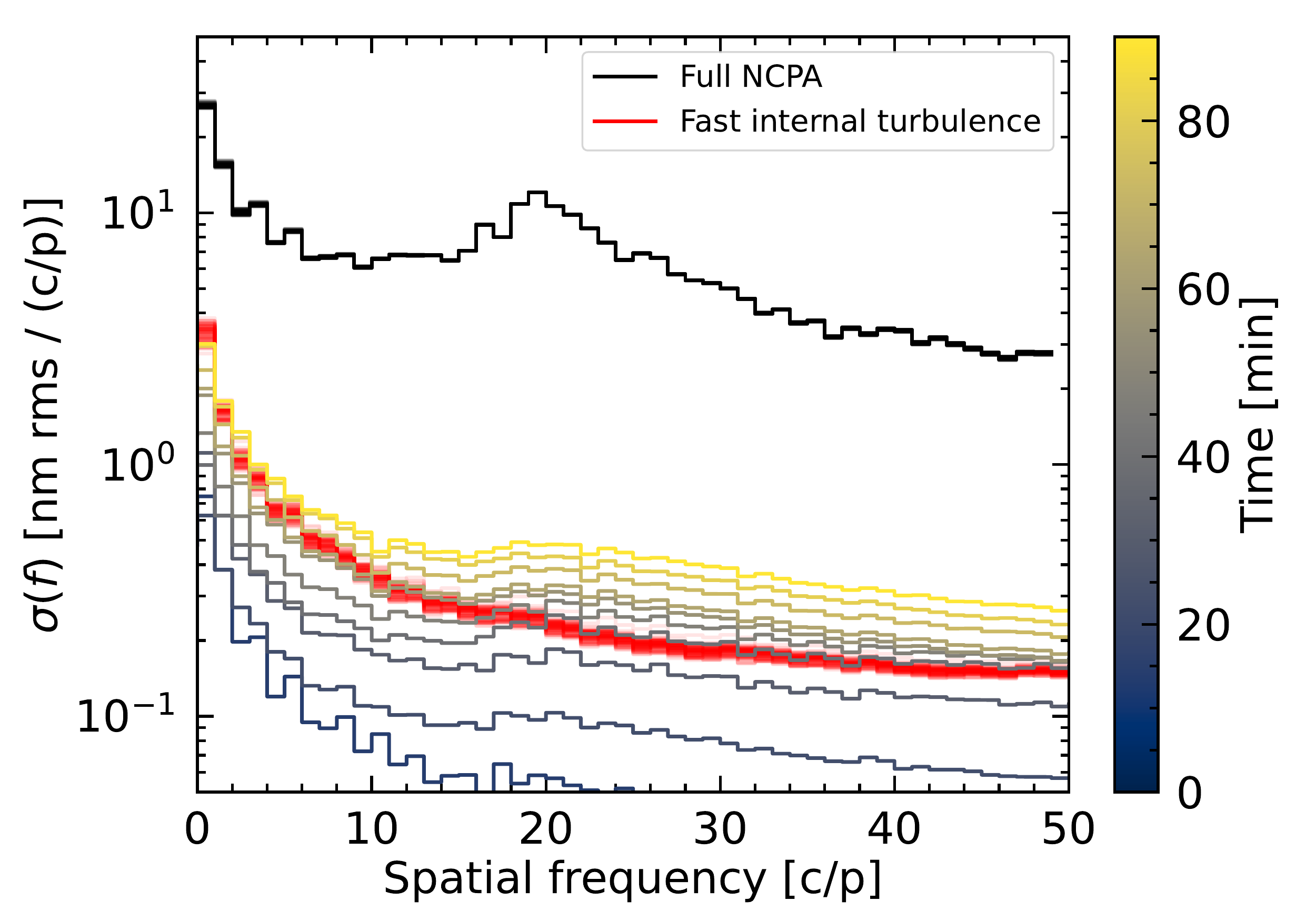}
  \caption{Relative contributions of the full NCPA (black), the estimated fast internal turbulence (red), and the slow-varying NCPA at different times (increasing with time, from blue to yellow), measured on the 2017-03-25\,b sequence. The curves correspond to averages, in consecutive bins of 3\,min for the full NCPA and the fast internal turbulence and 10\,min for the slow-varying NCPA.}
  \label{fig:temporal_aberrations_budget}
\end{figure}

Another way to look at the data is to compute $\sigma(f)$ for different components of the NCPA. In Fig.~\ref{fig:temporal_aberrations_budget} we show the values of $\sigma(f)$ computed on the 2017-03-25\,b sequence for the full NCPA, the estimated fast internal turbulence and the estimated slow-varying NCPA. This plot shows that the fast internal turbulence and the slow-varying NCPA are at least one order of magnitude smaller than the full NCPA, which is dominated by a static part originating in the various aberrations of the optics of the instrument. The contribution of the fast internal turbulence does not vary significantly with time, while the slow-varying NCPA slowly increases. This is expected because the beginning of the sequence is used as the reference to monitor the slow variations in the rest of the sequence. However, the two contributions end up at similar levels on timescales of several dozens of minutes at low spatial frequencies, and the slow-varying NCPA even dominate by a factor $\sim$2 beyond 20\,\cpup. This means that the two terms can potentially have an equivalent impact on the error budget of the instrument, and on its final contrast performance. 

The exact physical cause for the rise in NCPA beyond a few \cpup is difficult to pinpoint precisely. In SPHERE there are more than 15 differential optical elements (and between two and three times as many surfaces) between the point where the visible and NIR infrared beams split and the points where the wavefronts are measured in both channels, with the ExAO Shack-Hartmann and with ZELDA, respectively. With that many elements, even minor thermal variations can induce small misregistrations between optical elements and result in wavefront variations in the intermediate or high spatial frequences. The particular case of the ADCs is discussed in later sections.

\subsection{Impact on coronagraphic and differential images}
\label{sec:impact_coronagraphic_images_temporal}

\begin{figure*}
  \centering
  \includegraphics[width=1.0\textwidth]{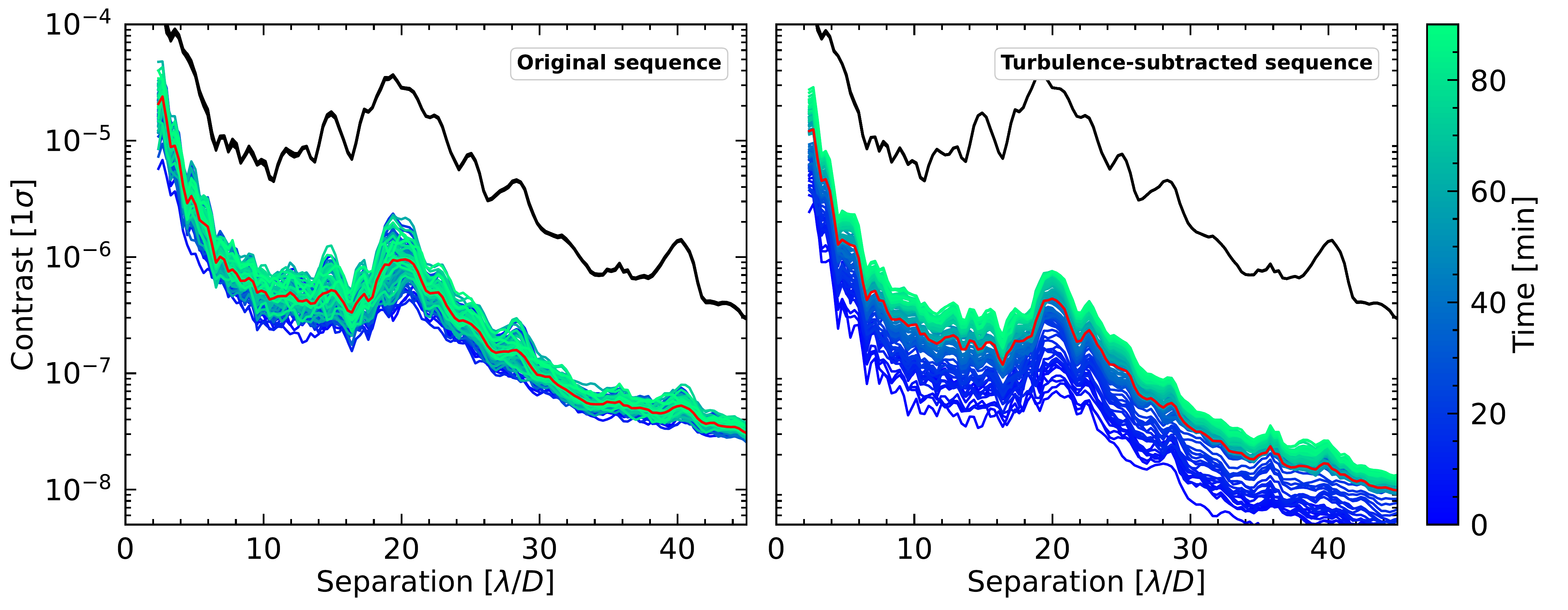}
  \caption{Sensitivity at 1$\sigma$ in direct coronagraphic images (black lines) and in speckle-subtracted images (colored lines) as a function of time from 0 to 90 minutes. The profiles are obtained from simulated coronagraphic images based on the 2017-03-25\,b original sequence of OPD maps (left) and on the turbulence-subtracted sequence (right). The red curve corresponds to the average sensitivity in coronagraphic images over the full duration of the sequence.}
  \label{fig:temporal_contrast}
\end{figure*}

\begin{figure*}
  \centering
  \includegraphics[width=1.0\textwidth]{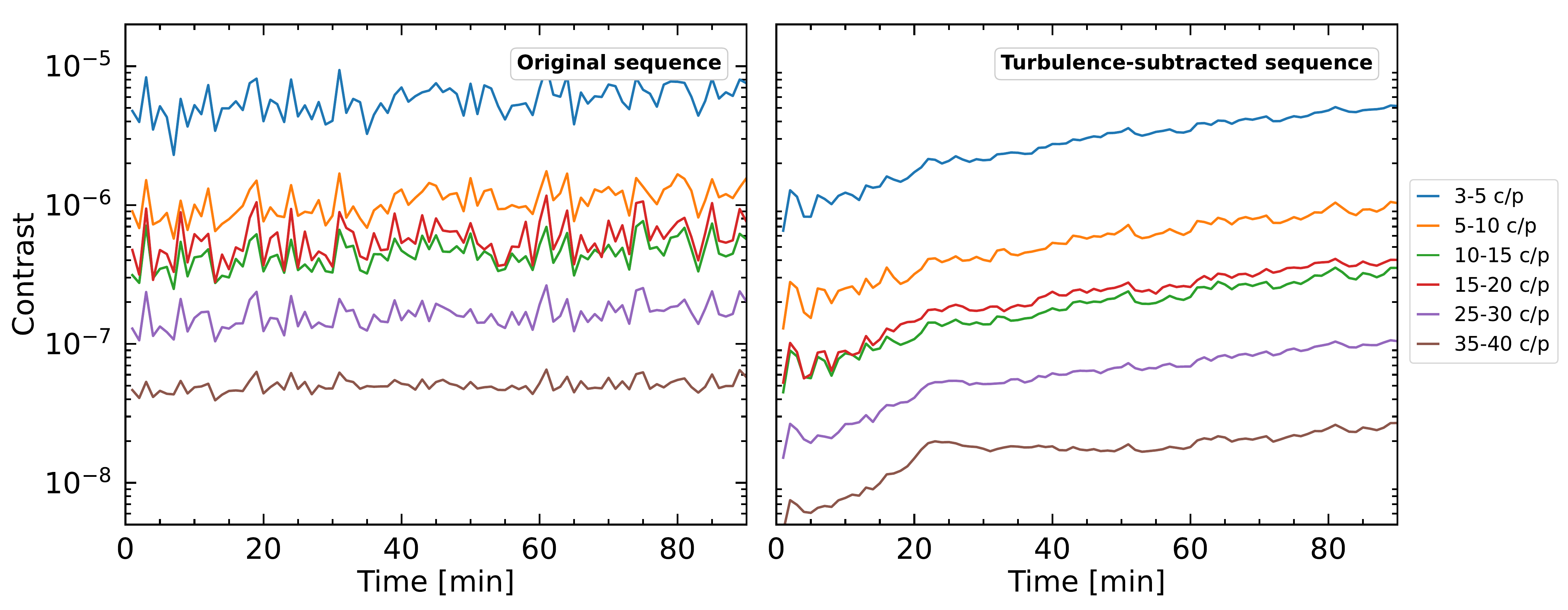}
  \caption{Sensitivity at 1$\sigma$ in speckle-subtracted images as a function of time for different bins of angular separation. The input profiles are those presented in Fig.~\ref{fig:temporal_contrast} and correspond to the 2017-03-25\,b sequence.}
  \label{fig:temporal_contrast_time}
\end{figure*}

The temporal decorrelation of NCPA has a direct impact on performance because of the decorrelation of the speckles that it induces in focal-plane coronagraphic images. All modern post-processing techniques rely on a relative stability of the speckles with time, on either short (seconds), moderate (hours), long (days), or very long (months to years) timescales. Any decorrelation of the speckles will therefore impair the ability to calibrate and subtract them.

To investigate the implications of our findings on the temporal evolution of the NCPA, we use the coronagraphic image reconstruction model developed in \citetalias{Vigan2019}. Briefly, this model uses various inputs such as amplitude maps (pupil amplitude errors, apodizer, Lyot stop) or phase maps (NCPA, apodizer substrate errors) in a simplified coronagraphic image formation model. The full model and its inputs are described in Appendix C of \citetalias{Vigan2019}. This model was found to provide an excellent match with actual images acquired on the internal source.

We use this model to simulate coronagraphic images at various time intervals from 0 to 90 minutes. For the NCPA, we use either the original sequences or the turbulence-subtracted sequences. Since we want to study the fundamental limits induced by the variations of the NCPA, the simulated images do not include any detection noise. Then, we perform a simple post-processing procedure where the first image of the sequence is subtracted from all subsequent images in order to quantify the impact of the NCPA evolution on the correlation of the speckles. The final sensitivity in the residual images is estimated by computing the azimuthal standard deviation as a function of angular separation. Results for the 2017-03-25\,b sequence are presented in Fig.~\ref{fig:temporal_contrast}.

In raw coronagraphic images the fast internal turbulence and the slow-varying NCPA have a negligible impact, as can be expected for aberrations at a level of a few nanometers compared to a total close to 60\,nm\,rms. The 1$\sigma$ sensitivity in raw coronagraphic images is around $10^{-5}$ between 6 and 14\,\lsd, and the spread with time is negligible, which indicates that the raw sensitivity is dominated by the static part of the NCPA.

However, the situation is different when looking at the sensitivity in speckle-subtracted images. In the original sequence, the profiles appear chaotic around an average profile that is about one order of magnitude lower than the sensitivity in raw images, and there is no clear trend with time. This means that the speckles are slightly decorrelated from one image to the next, which is highly consistent with an internal turbulence that varies on timescales on the order of a few seconds.

The profiles for the turbulence-subtracted sequence show two major differences: (1)~there is a clear trend with time, and (2) the contrast level is at least a factor of 2 better than with the original sequence. Since the fast internal turbulence has been subtracted to a very low level, the dominating effect becomes the slow linear decorrelation of the NCPA described in Sect.~\ref{sec:slow_linear_decorrelation}, which acts over timescales of several minutes. The speckles in consecutive images are therefore highly correlated, leading to an improved contrast in speckle-subtracted images. As the time difference increases between images, the slow decorrelation of the NCPA induces small changes in focal-plane speckles and leads to a less efficient subtraction. However, and contrary to the effect of the fast internal turbulence, the decorrelation is slow and continuous, so the loss in contrast appears smooth and the trend with time becomes visible.

Another way to look at the effect of the fast internal turbulence is provided in Fig.~\ref{fig:temporal_contrast_time}, which shows the average 1$\sigma$ sensitivity over different ranges of angular separations as a function of time. In the original sequence the sensitivity remains almost flat on average with time, but with a chaotic behavior around the average at each separation that corresponds to the random effect of the fast internal turbulence. The chaotic behavior induces variations in contrast of a factor 2 or less from one image to the next. In the turbulence-subtracted sequence, the variations in contrast are much less chaotic and there is a clear trend of decreasing contrast with time. And very importantly, the average contrast at the same separations in both sequences is a factor 4 to 5 better in favor of the turbulence-subtracted sequence.

\begin{figure}
  \centering
  \includegraphics[width=0.49\textwidth]{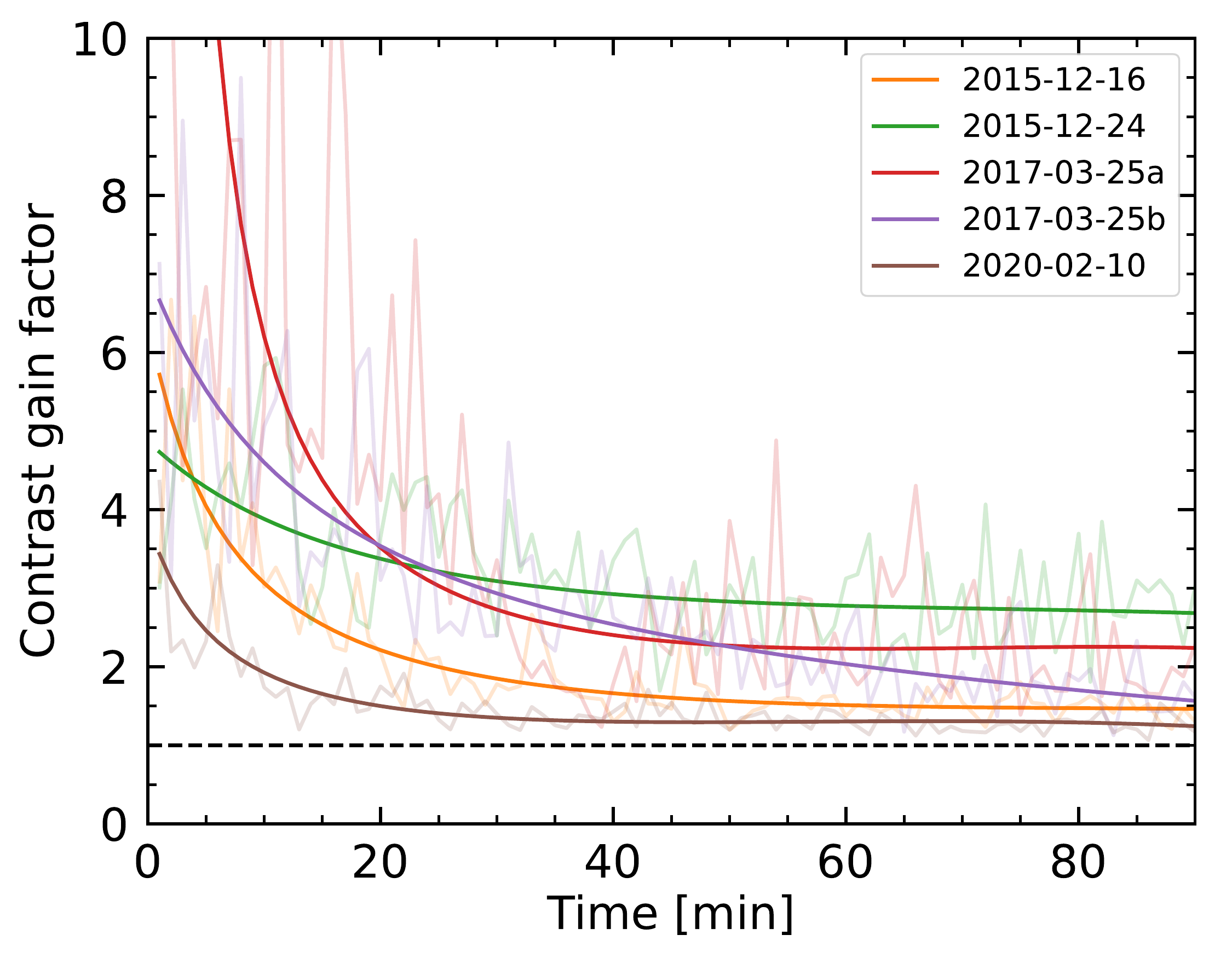}
  \caption{Average gain in sensitivity in speckle-subtracted images as a function of time between the sequences where the fast internal turbulence has been subtracted and the original ones. The gain in contrast is averaged between 10 and 15\,\lsd. The raw data are displayed in lighter colors, while the low-order polynomial fits are displayed in bright colors to improve readability. The gain in contrast is always greater than unity, which means that the sensitivity is always better in the turbulence-subtracted sequences.}
  \label{fig:temporal_contrast_gain_time}
\end{figure}

Finally, we confirm that the fast internal turbulence is a dominating effect in all sequences. In Fig.~\ref{fig:temporal_contrast_gain_time} we show the gain in contrast between 10 and 15\,\lsd as a function of time for all sequences. The data is fitted with a low-order polynomial to smooth out the chaotic variations induced by the original data. Although a large variability is observed between the sequences, the gain in sensitivity is always greater than unity, which means that the sensitivity is always better in the turbulence-subtracted sequences. In particular, the gain is higher than 2 in all sequences for timescales shorter than 10\,min, which is to be compared to the typical 5\,min frequency used in the star-hopping observing strategy \citep{Wahhaj2021}. Minimizing the fast internal turbulence directly in the hardware could therefore potentially lead to contrast improvements with that specific observing strategy.

Based on our analysis, we conclude that the fast internal turbulence is currently the limiting factor of the ultimate contrast performance in SPHERE. The variability observed between the sequences certainly deserves deeper investigations, in particular to understand its relation with respect to temperature, pressure, or other external parameters. This will be further studied in the future when trying to confirm the origin of the internal turbulence and attempting to decrease its impact.

\section{Impact of the derotator}
\label{sec:impact_derotator}

The derotator is a critical optical element of SPHERE, located very early in the optical train. Because SPHERE is installed on the Nasmyth platform of the VLT-UT3, which is an alt-az telescope, the derotator is mandatory to stabilize either the pupil or the field with respect to the instrument. For high-contrast observations, which are typically obtained to allow the use of ADI, the derotator is operated in what is known as the pupil-stabilized (or pupil-tracking) mode. There are two advantages in keeping the pupil fixed with respect to the instrument. The first is that the spider vanes supporting the secondary mirror remain fixed with respect to the optics, which allows implementing a well-optimized Lyot stop for the coronagraph and produces a better attenuation of the diffraction. The second is that the induced field rotation generates diversity in the data, which is used a posteriori by various speckle-subtraction techniques.

Like any optical element in the beam, the derotator will introduce some aberrations that may vary depending on its orientation. This is true for any optical element located upstream of the derotator in the optical train. In this section we study the impact of the aberrations introduced by the motion of the derotator when used in pupil-stabilized mode.

\subsection{Experimental data}
\label{sec:experimental_data_derot}

On the Nasmyth platform of an alt-az telescope, stabilizing the pupil only requires us to compensate for the change in elevation of the target\footnote{\url{http://www.eso.org/sci/libraries/historicaldocuments/VLT_Reports/VLTrep63_A1b.pdf}}. Considering observations of at most 1.5 hours symmetric around the meridian (i.e., at hour angles varying from $-0.75$\,hr to $+0.75$\,hr), the variation in altitude for targets observable from the VLT\footnote{The VLT has an exclusion region of 3\degre around zenith.} will reach a maximum of $\sim$9\degre. Due to the implementation of the derotator using a K-mirror system, a rotation of $\theta$ degrees of the pupil only requires a physical rotation by $\theta/2$ of the derotator. For targets observable from the VLT this means that the derotator will move by at most $\sim$4.5\degre. Moreover, because the altitude reaches a maximum when the target crosses the meridian and then decreases again, the derotator will rotate in one direction until meridian passage and will change direction afterward. This is a particularly interesting property from the optical aberrations point of view because the beam will see the same aberrations for symmetric positions around the meridian (i.e., at hour angles of $-h$ and $+h$).

On 2020 February 29 we acquired a ZELDA sequence to simulate the derotator motion for a target with declination $\delta = -15\degree$ observed for 1.5\,hr around meridian passage. This corresponds to a variation of $\pm$4.664\degree\ in altitude, and therefore a motion of $\pm$2.332\degree\ for the derotator. Instead of simulating a continuous motion of the derotator, we moved the derotator at 50 discrete positions linearly spaced in hour angle. Two series of data were acquired: one with the ZELDA mask and one with the clear pupil, as explained in Sect.~\ref{sec:experimental_data_temporal}. In order to minimize the impact of the fast internal turbulence, which was already identified and known to have a characteristic timescale of a few seconds, we introduced a neutral density in the optical path to be able to use a DIT of 20\,s and average out a significant fraction of the internal turbulence. Longer DITs would have been even more appropriate, but the test was performed in a time-constrained context. The result is a sequence where the fast internal turbulence is attenuated, but not completely removed. The sequence of ZELDA images was acquired over a total duration of 27\,min.

The OPD maps corresponding to each derotator position were produced using \texttt{pyZELDA} and are the starting point of the analysis presented below.

\subsection{Analysis}
\label{sec:analysis_derot}

\begin{figure}
  \centering
  \includegraphics[width=0.4\textwidth]{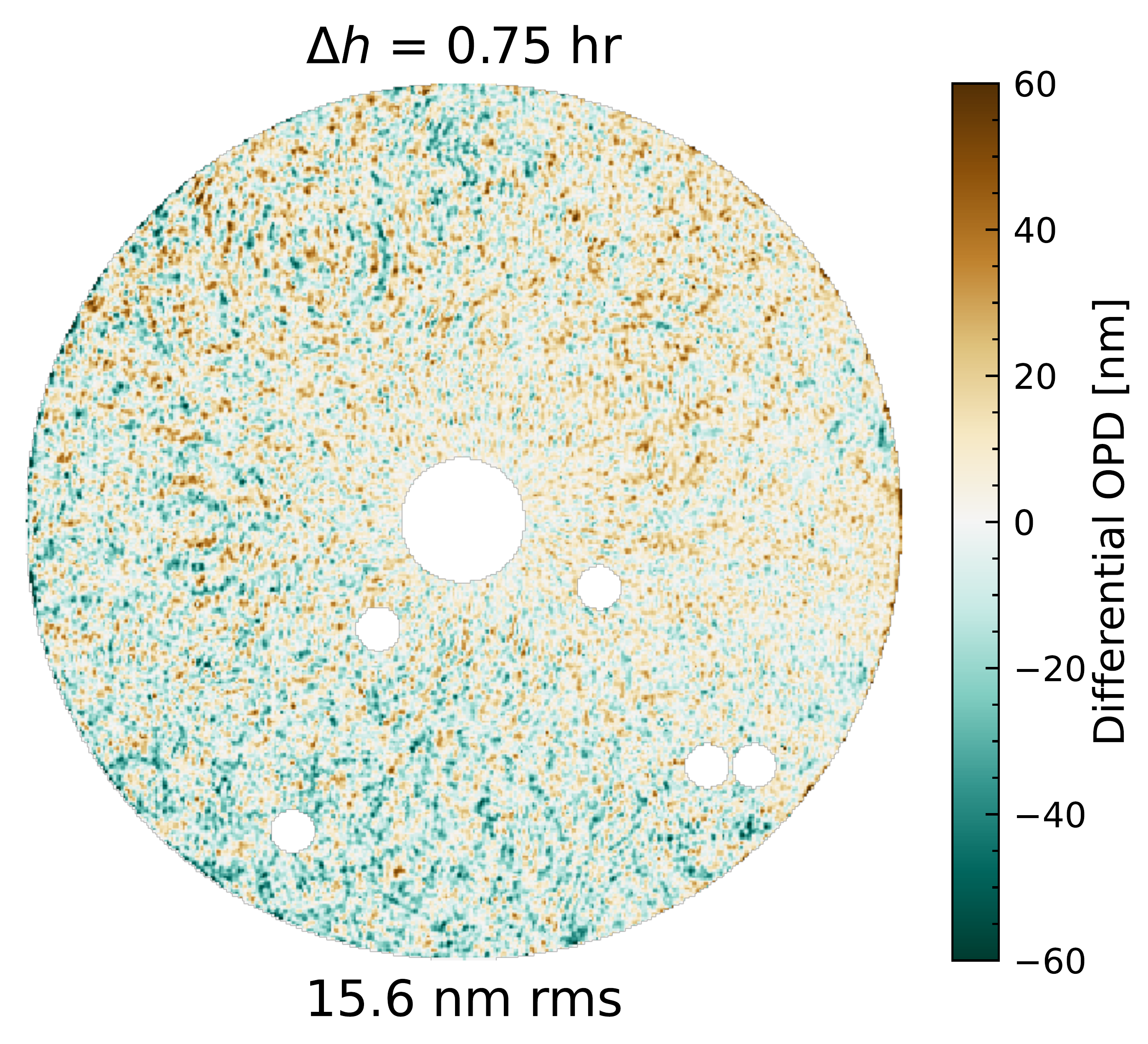}
  \caption{Residual aberrations when subtracting two OPD maps acquired at derotator positions corresponding to a target at declination $\delta = -15$\degre observed at hour angles of $0$\,hr and $+0.75$\,hr. The dead actuators of the deformable mirror and the region corresponding to the central obscuration have been masked. The standard deviation of the residuals is reported below the residual map.}
  \label{fig:derotator_residuals}
\end{figure}

\begin{figure*}
  \centering
  \includegraphics[width=1\textwidth]{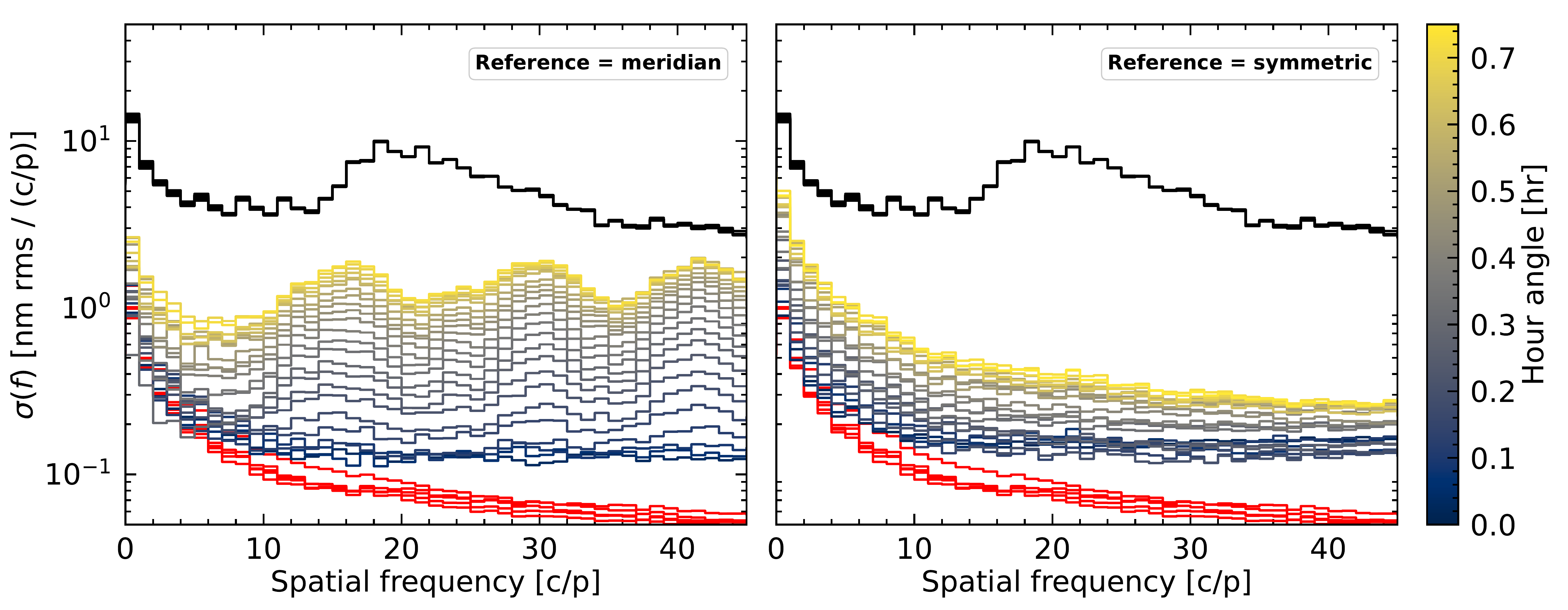}
  \caption{Residual aberrations in differential OPD maps obtained for different orientations of the derotator, simulating the observation of a target at declination $\delta = -15$\degre and at hour angles from $-0.75$\,hr to $+0.75$\,hr. The black curves corresponds to $\sigma(f)$ for OPD maps before any subtraction. In the left plot, the OPD map obtained at meridian passage is used as reference and subtracted from all subsequent maps. The $\sigma(f)$ for the residuals when subtracting the reference map from the preceding maps is not shown because it is almost identical. In the right plot, the OPD maps are subtracted in pairs (two at a time) symmetrically around meridian passage (i.e., for almost identical orientations of the derotator). The red curves show the contribution for the estimated fast internal turbulence over 20\,s timescales for the temporal sequences presented in Sect.~\ref{sec:temporal_decorrelation}.}
  \label{fig:derotator_aberrations}
\end{figure*}

\begin{figure*}
  \centering
  \includegraphics[width=1\textwidth]{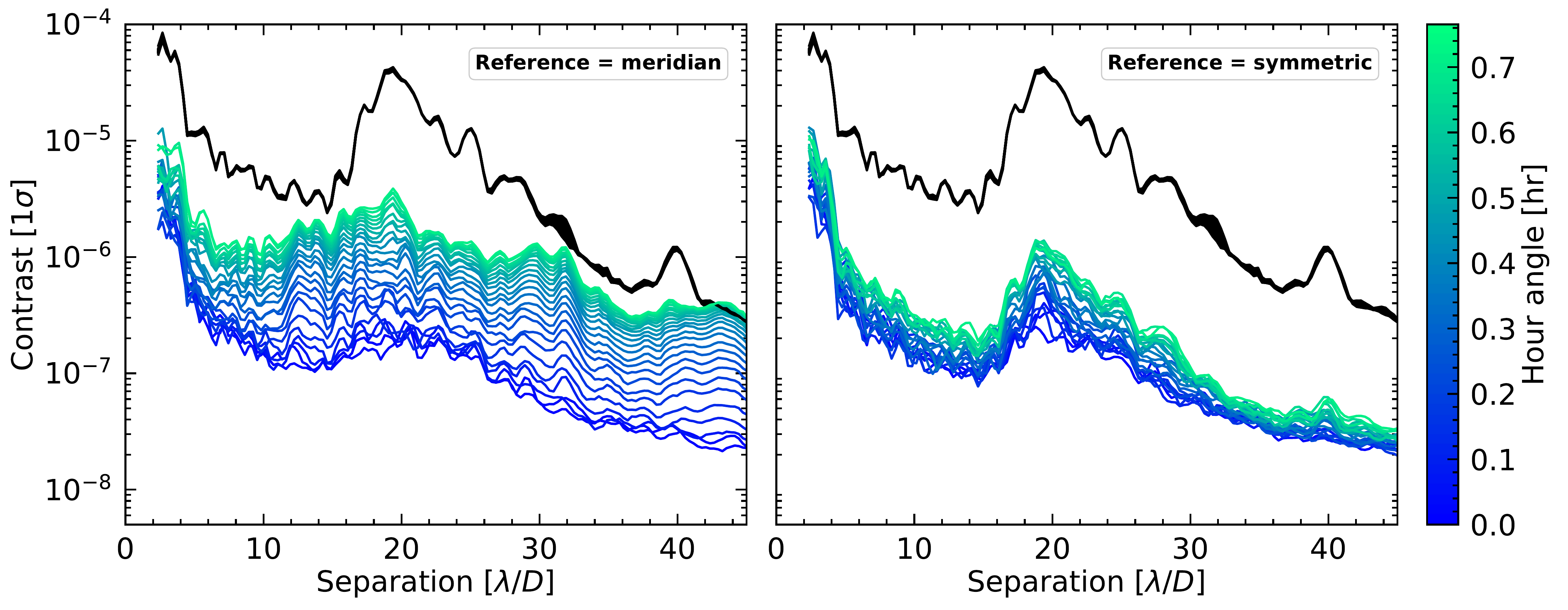}
  \caption{Sensitivity at 1$\sigma$ in direct coronagraphic images (black lines) and in speckle-subtracted images (colored lines) as a function of hour angle from 0 to 0.75\,hr. The speckle subtraction strategy follows the strategies described for the OPD maps in Sect.~\ref{sec:analysis_derot}. In the left plot, the image acquired closest to meridian passage is subtracted from all subsequent (or preceding) images, while in the right plot the pairs of images are acquired at opposite hour angles with respect to meridian passage.}
  \label{fig:derotator_contrast}
\end{figure*}

We again base our analysis on the differences between pairs of OPD maps. We first use the OPD map acquired at the equivalent of meridian passage as the reference and we subtract it from all of the other OPD maps. The residuals obtained for the OPD map obtained at hour angle $h = +0.75$\,hr are presented in Fig.~\ref{fig:derotator_residuals}. They show a mix of high spatial frequency patterns closer to the center and lower spatial frequency patterns closer to the edge of the pupil. A very low spatial frequency aberration, dominated by tip and tilt along the diagonal from bottom left to top right, is also visible across the whole pupil.

The level of residuals $\sigma(f)$ is presented in Fig.~\ref{fig:derotator_aberrations} (left). The amount of residuals with increasing hour angle, that is with increasing derotator angle, is color-coded and shows a clear trend with time. The residuals presented in Fig.~\ref{fig:derotator_residuals} correspond to the yellow curve for $\Delta h = +0.75$\,hr. For small derotator angle differences the level of residuals appears flat and at a level of $\sim$0.15\,nm\,rms down to $\sim$10\,\cpup. There is a steep increase below 10\,\cpup, but it is due in large part to the residual fast internal turbulence that has not been completely averaged over DITs of 20\,s (see next paragraph). The derotator is seen by the WFS of the SPHERE ExAO system, so the very low spatial frequencies that it introduces when it rotates should be corrected. Beyond 10\,\cpup, the levels of residuals slowly rise due to the mismatch between aberrations at different orientations. The exact origin of the wiggles that appear in $\sigma(f)$ is not precisely understood and would require further investigations. Ignoring what happens at very low spatial frequencies, the amount of residuals increases by approximately one order of magnitude over the range of considered hour angles.

The contribution of the residual fast internal turbulence over 20\,s exposures is estimated using the temporal sequences from Sect.~\ref{sec:temporal_decorrelation}. The raw images of the sequences are first temporally binned to simulate exposures of $\sim$20\,s. The OPD maps are then recomputed, and the fast internal turbulence is estimated in the same manner as explained in Sect.~\ref{sec:subtraction_internal_turbulence}. Finally, we compute the average $\sigma(f)$ for the fast internal turbulence for each sequence and we overplot them in red in Fig.~\ref{fig:derotator_aberrations}. Although we are comparing different dates, the amplitude of the fast internal turbulence appears similar in our different sequences, so we assume that the fast internal turbulence on 2020 February 29 would be of the same order of magnitude. With this assumption, the rise of $\sigma(f)$ for differential OPD maps below 10\,\cpup is perfectly consistent with an effect of the residual fast internal turbulence.

We also consider a second subtraction strategy where we subtract OPD maps acquired for positions of the derotator symmetric with respect to meridian passage. If everything remains static, the subtraction should be close to perfect because the derotator should be exactly in the same orientation for opposite hour angle values. Of course, this is not exactly the case because of the residual fast internal turbulence and the slow-varying NCPA, and because the acquisition of the frames will never be exactly symmetric with respect to meridian passage. The level of residuals for this strategy are presented in Fig.~\ref{fig:derotator_aberrations} (right).

The difference with the first strategy is striking: the residuals are significantly lower beyond a few \cpup and remain almost flat in the spatial frequencies not affected by the residual internal turbulence. In the present case, almost the same aberrations are seen by the beam for symmetric positions across the meridian, which induces a much lower level of residuals in differences of OPD maps. This clearly confirms the importance of observations performed symmetrically across the meridian. Of course, this result was to be expected by design, but to the best of our knowledge this is the first experimental demonstration of the importance of this observing strategy. Major recent direct imaging surveys like SHINE or GPIES based their observing strategies on ADI and tried to optimize their night schedules to observe targets across meridian passage to benefit from the highest possible field-of-view rotation \citep[e.g.,][]{Lagrange2016}. The symmetry of the observations with respect to meridian passage is never exact, however. The atmospheric residuals from the ExAO are, in general, the limiting factor, so the impact of our findings on the overall astrophysical results of such surveys is probably limited. The impact could potentially become noticeable for observations in very stable conditions where the quasi-static speckles become the limiting factor.

We observe a higher level of residuals in difference of OPD maps taken farther apart in time but the increase is flat and relatively moderate, with only a rise from 0.15 to 0.45\,nm\,rms at 20\,\cpup and from 0.15 to 0.6\,nm\,rms at 10\,\cpup. This increase is slightly larger than what could be expected from the slow NCPA decorrelation over 27\,min, but only by a factor of 2 or 3. The low spatial frequencies are highly contaminated by the fast internal turbulence, but at higher frequencies a combination of the unaveraged internal turbulence and the slight mismatch in derotator position could easily account for the remainder.

Our analysis of the residuals following two distinct strategies confirms the importance of observations performed across the meridian. Highly asymmetric observations across the meridian would resemble the extreme case presented in the first subtraction strategy, which would have a major impact on the final contrast. On the contrary, very symmetric observations should mainly be impacted by the slow variations of the NCPA. Both strategies indicate that observations taken close in time and in derotator positions have a lower level of residuals. However, it may not always be possible to subtract images acquired close in time due to self-subtraction effects on any putative companion in the data if the field rotation was not sufficient \citep[see, e.g.,][]{Pueyo2016}. In the second strategy, symmetric images with respect to the meridian are mostly distant in time, which would reduce self-subtraction effects except for images acquired very close to meridian passage.

\subsection{Impact on coronagraphic and differential images}
\label{sec:impact_coronagraphic_images_derot}

We simulate coronagraphic images in the same way as described in Sect.~\ref{sec:impact_coronagraphic_images_temporal}. The estimation of the final sensitivity is done by subtracting pairs of images following the two strategies presented in the previous section and by computing the azimuthal standard deviation in the residual images. The results are presented in Fig.~\ref{fig:derotator_contrast}.

As expected, the final sensitivity is much worse when following the first subtraction strategy. In both cases, subtracting two images acquired close in time results in a sensitivity of $\sim$$10^{-7}$ at 15\,\lsd. This is in line with the results presented in Sect.~\ref{sec:impact_coronagraphic_images_temporal} in the turbulence-subtracted sequences. However, with the first strategy there is a loss of more than an order of magnitude at some separations when subtracting the reference image at meridian passage from the one obtained 0.75\,hr before or after. The loss of contrast in the second strategy is of a factor of $\sim$2, even when subtracting images acquired up to 1.5\,hr apart. This loss of sensitivity is compatible with the loss expected from the slow decorrelation of NCPA.

\section{Realistic observing sequence}
\label{sec:realistic_observing_sequence}

\begin{figure}
  \centering
  \includegraphics[width=0.49\textwidth]{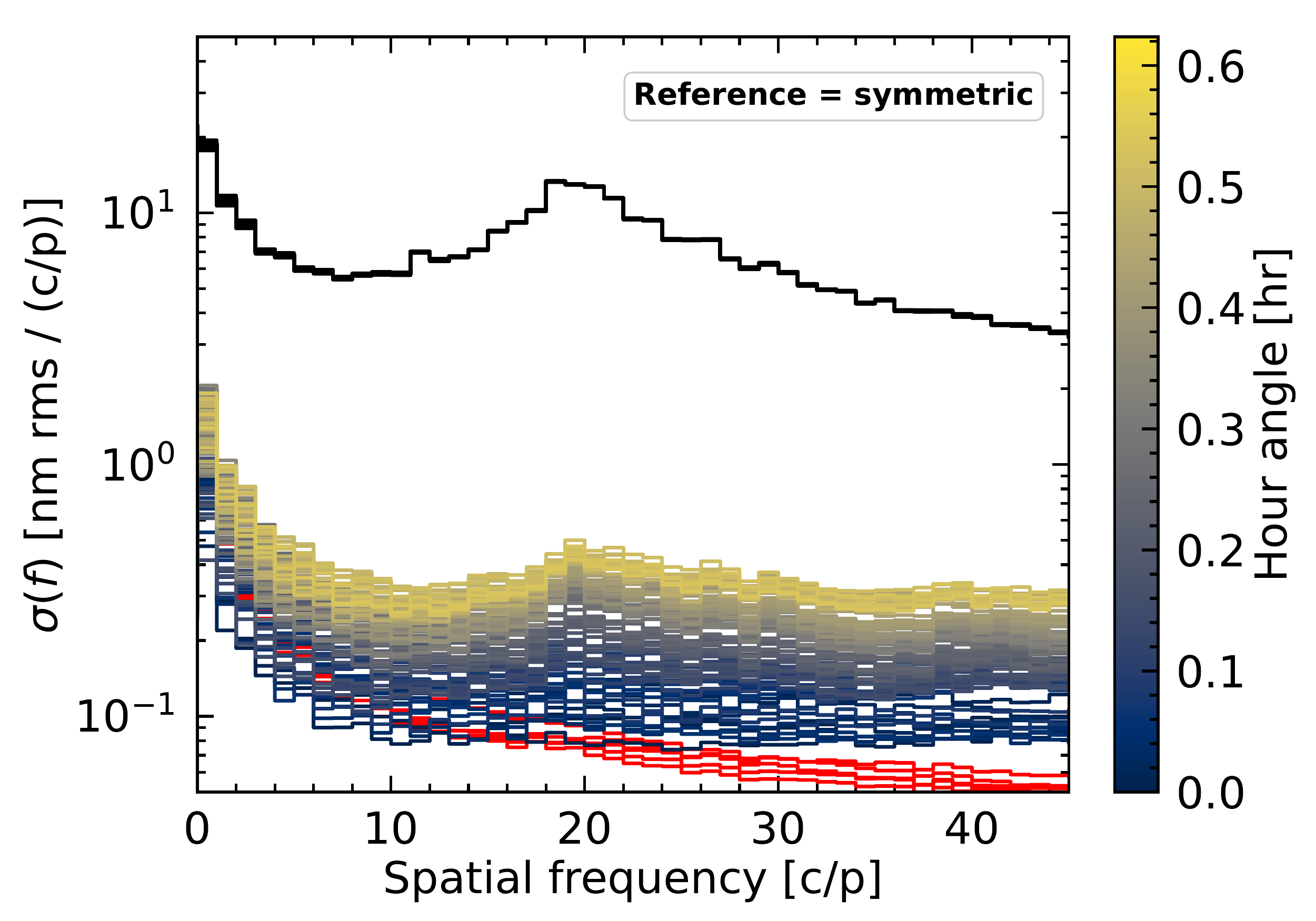}
  \caption{Residual aberrations in differential OPD maps obtained for the simulated realistic observation of a target at declination $\delta = -37$\degre and at hour angles from $-0.62$\,hr to $+0.53$\,hr. In this sequence the UT3 telescope was tracking inside the dome, and the derotator and ADCs were tracking as in a real observation. The OPD maps are subtracted by quasi-symmetric pairs around meridian passage, that is for almost identical orientations of the derotator. The red curves show the contribution for the estimated fast internal turbulence over 20\,s timescales for the temporal sequences presented in Sect.~\ref{sec:temporal_decorrelation}.}
  \label{fig:telescope_aberrations}
\end{figure}

The final possible contributor to the wavefront error budget during actual observations are the ADCs in the optical train of the instrument. SPHERE has two sets of ADCs, one for the visible path that includes the ExAO wavefront sensor and the scientific instrument ZIMPOL and one for the NIR path that includes the two scientific instruments IRDIS and IFS. The ADCs are located after the visible-NIR dichroic, which means that they can potentially introduce differential wavefront aberrations between the two beams.

To assess the impact of the ADCs, a sequence of ZELDA OPD maps was acquired on the internal source during daytime on 2015 December 24 with the UT3 tracking inside the dome for $\sim$1\,hr, as if it were pointing at a target at a declination $\delta = -37$\degre. The data was acquired for hour angles equivalent to $h = -0.62$ to $+0.53$\,hr. From the point of view of the instrument, this sequence is equivalent to on-sky observations except for the presence of atmospheric disturbances, which means that both the derotator and the ADCs were tracking continuously as in normal observations. The sequence was repeated twice, once to acquire a sequence of images with the ZELDA mask in the beam and once to acquire a sequence of clear pupil images. A DIT of 1\,s was used during the whole sequence to provide a high cadence of observations.

For the analysis of the data we binned the images to equivalent DITs of 20\,s to minimize the impact of the fast internal turbulence. This was done a posteriori because the fast internal turbulence had not yet been identified in 2015 when the sequence was acquired. Each ZELDA pupil image was associated with the clear pupil image with the closest telescope altitude value, which is equivalent to the closest derotator orientation. The OPD maps were then reconstructed with \texttt{pyZELDA} as in Sects.~\ref{sec:temporal_variation} and~\ref{sec:impact_derotator}.

We performed again subtractions of OPD maps following the second strategy described in Sect.~\ref{sec:analysis_derot} where pairs of OPD maps are selected to be as closely symmetric as possible in altitude before and after the meridian. The results in terms of $\sigma(f)$ are presented in Fig.~\ref{fig:telescope_aberrations}. Qualitatively, these results are similar to those obtained with the derotator-only test: at low spatial frequencies the differences are likely dominated by the residual fast internal turbulence on timescales of 20\,s, and beyond 10-15\,\cpup there is a slow increase in the residuals with time. The increase in $\sigma(f)$ over the 1\,hr test is on the order of a factor of 4. It was smaller than a factor of 2 for the derotator-only test. That test was half as long, but the increase was already greater than  expected from only the slow-varying NCPA. It is therefore tempting to attribute the observed increase in the present analysis to the combined effect of the slow-varying NCPA, the mismatch in derotator orientation, and the ADCs. Unfortunately, it is impossible to easily evaluate the respective contribution of each of these terms. A dedicated test where only the ADCs are tracking while the derotator remains static would be necessary for a more detailed investigation. However, it does show that the residual aberrations in differences of OPD maps, when following the proper observing strategy, can reach very low levels, which we translate into highly correlated speckles in the coronagraphic images.

\section{Conclusions and discussion}
\label{sec:conclusions_discussion}

Non-common path aberations are a dominating effect in coronagraphic images produced by new-generation high-contrast imagers like SPHERE, GPI, or SCExAO. These instruments have been optimized at every stage to minimize the aberrations and maintain their variations at an extremely low level; nonetheless, thermal and mechanical changes will induce small variations over timescales on the order of minutes to hours. In the present work we studied these variations based on ZELDA, the Zernike wavefront sensor embedded in VLT/SPHERE. Contrary to previous studies of the NCPA evolution based on the decorrelation of speckles in focal-plane images, we were able to directly measure the wavefront at the level of the coronagraph and precisely study the NCPA thanks to long sequences of measurements on the internal source. Our results are specific to the VLT/SPHERE instrument, but they contain some interesting lessons for future high-contrast imaging instruments, for example on extremely large telescopes that will target the $10^{-8}$--$10^{-9}$ contrasts \citep[e.g.,][]{Kasper2021}.

Our biggest finding, which was unexpected, is the presence of a fast internal turbulence at a level of a few nanometers rms at low spatial frequencies and with a characteristic timescale of a few seconds. This seems small at first look compared to a static error budget of $\sim$60\,nm\,rms, but it becomes a dominating term when performing subtraction of images, which is the basis of almost all current speckle-subtraction post-processing techniques. Even if this turbulence tends to average over long exposures, it could become a limiting factor for bright stars where the ExAO performance, and therefore the final contrast performance, are expected to be the highest when observing conditions are optimal. For example, with SPHERE/IRDIS bright stars up to $H = 1 - 2$ are routinely observed with DITs below a few seconds. We have also seen that the impact of the turbulence is still visible on much longer timescales, up to 20\,s or more, which concerns observations of fainter targets.

The fast internal turbulence is likely created by a motor placed underneath the NIR beam and which is at a temperature a few degrees above that of the enclosure. In retrospect, this seems to be an unfortunate design choice that could easily be avoided in future instruments. It also acts as a warning that when aiming for extremely high-contrast performance with extremely low levels of aberrations, every detail counts. It is also important to understand the origin of this fast internal turbulence in the context of ongoing or foreseen evolutions of SPHERE, such as the coupling with the high-resolution spectrograph CRIRES+ \citep[HiRISE;][]{Vigan2018hirise,Otten2021}, or the implementation of a fast (3\,kHz) second-stage ExAO system and additional improvements \citep{Boccaletti2020}.

Some of the key advantages of ZELDA are its flexibility, its simple implementation, and the fact that it can easily measure a wide range of spatial frequencies \citep[e.g.,][]{Pourcelot2021}. The presence of this sensor inside SPHERE was not originally planned and is the result of several happy coincidences \citep{N'Diaye2014,N'Diaye2016spie}. It has nonetheless revealed to be an amazing diagnostic tool that enabled the development of a deep understanding of SPHERE and revealed some unexpected effects such as a derotator mechanical backlash \citep{Beuzit2019} or the infamous low-wind effect \citep{Sauvage2015}. This type of monitoring capability is extremely important for future exoplanet imagers and is already considered for both monitoring and compensation of the NCPA in future instruments, either in space \citep{Shi2016} or on ELTs \citep{Carlotti2018a} as a way to improve direct exoplanet detection \citep[e.g.,][]{Houlle2021}. It also highlights the importance of closely monitoring high-contrast imaging instruments, from the point of view of the aberrations, or of the temperature or vibrations in some key locations. One could also foresee monitoring the turbulence inside of the instrument's enclosure \citep[e.g.,][]{Ziad2013} if these effects could potentially affect the performance.

When subtracting the contribution of the fast internal turbulence, we observe a linear evolution of the NCPA that is consistent with expectations for high-contrast imagers \citep[e.g.,][]{Martinez2012} on the order of fractions of picometers per second. Again, while this may seem small on the timescale of seconds, it integrates to a few nanometers over the timescales relevant for typical astrophysical observations. When considering differences of coronagraphic images on the internal source, this can result in a loss of contrast of up to one order of magnitude, from $10^{-7}$ up to $10^{-6}$.

The amplitude of the slow-varying NCPA over several dozens of minutes is similar to the amplitude of the fast internal turbulence, which means that the two terms have an equivalent effect on the ultimate performance. Even though they have very different timescales, they can both impact the level of residuals in subtraction of coronagraphic images. The fast internal turbulence affects all possible subtractions, while the slow-varying NCPA mostly affects subtraction of images very distant in time. The latter is a potential issue as observations performed symmetrically across meridian passage of the targets are the most efficient in removing the speckles induced by the derotator. Our measurements with ZELDA confirm that the impact can be significant on the contrast curve when considering the subtraction of images acquired for very different orientations of the derotator. However, and contrary to the fast internal turbulence, slow-varying NCPA could be calibrated regularly either using ZELDA or active wavefront control techniques \citep[e.g.,][]{Potier2020}.

Pushing the reasoning further, one would ideally avoid using a derotator altogether, as is possible for Cassegrain instruments like GPI, or performing observations at a fixed position of the device. For an instrument like SPHERE located on the Nasmyth platform of an alt-az telescope, this would result in a small rotation of the pupil, which would severely impact the efficiency of the Lyot stop. An interesting alternative could be to implement a rotating Lyot stop rather than a derotator. While the actual implementation could be extremely tricky if the Lyot stop is located in a cryostat, as in the case of SPHERE/IRDIS, some other implementation solutions could be foreseen from the start if this solution enables a significant gain in final performance. Another possibility could be the use of an array of micro-mirrors to create an adaptive Lyot stop, as has been proposed for adaptive pupil apodization \citep{Carlotti2018b}.

Our tests are quite representative of actual observations, but there is an important ingredient missing: a turbulent atmosphere between our point-like object and our detector. First, the atmospheric refraction induces a chromatic beam-shift, which means that the different wavelengths of the beam do not see the same parts of the optics located prior to the ADCs. This effect was of course considered in the SPHERE design, and the specifications of the optics was defined to make it acceptable to reach the $10^{-6}$ final contrast \citep{Beuzit2019}. One of the important requirements of SPHERE is to cover simultaneously a very broad band in the NIR due to the parallel observations with both IFS and IRDIS from $Y$ to $H$ band (and $K$ in some cases). However, this requirement can be relaxed if considering observations only in narrowband filters such as the dual-band imaging filters located in IRDIS \citep[$R \simeq 30$;][]{Vigan2010}, but remains an important factor in IFS observations where the use of the spectral information brings a massive gain in contrast \citep{Vigan2015,Mesa2017}. In any case, the chromatic beam-shift would also be affected by some variations over time due to global NCPA variations, so the final performance level including this term is difficult to evaluate without a thorough design study.

The turbulent part of the atmosphere, even filtered by a very efficient ExAO system, remains ultimately the main contributor of residual aberrations. In the SPHERE wavefront error budget, ExAO residuals are at a level of $\sim$65\,nm\,rms (for spatial frequencies below 20\,\cpup) on bright stars observed under median seeing conditions (typically 0.8\as). Whether or not the fast internal turbulence and the slow NCPA variations, which are at a level of a few nanometers at low spatial frequencies, can become important limitations of the final contrast performance is an open question. However, \citet{Milli2016} first discovered the fast decorrelation of the quasi-static speckles in data obtained on-sky, before reproducing the result on the internal source. This demonstrates that even small effects like this one can have a visible signature on top of atmospheric residuals.

Even if the effect is small with the current system, it may not remain that way with a very fast second-stage ExAO system such as the one considered for an upgrade of the SPHERE instrument \citep{Boccaletti2020}. Moreover, the effects that we observe are most prominent, and could become limiting, at small spatial frequencies (i.e., small angular separations) where a very fast second stage ExAO system would provide the highest gain and where new detections would be the most interesting from the astrophysical point of view. Our analysis is therefore an important step in the fine understanding of the system that is required for future upgrades and for the development of new exoplanet imagers that target very high contrasts in the visible.

\begin{acknowledgements}
  The authors are extremely grateful to ESO for granting them VLT technical daytime for these tests and to the Paranal observatory staff for their support. This project has received funding from the European Research Council (ERC) under the European Union's Horizon 2020 research and innovation programme (grant agreement No. 757561). This work was partly supported by the Action Spécifique Haute Résolution Angulaire (ASHRA) of CNRS/INSU co-funded by CNES. SPHERE is an instrument designed and built by a consortium consisting of IPAG (Grenoble, France), MPIA (Heidelberg, Germany), LAM (Marseille, France), LESIA (Paris, France), Laboratoire Lagrange (Nice, France), INAF - Osservatorio di Padova (Italy), Observatoire de Gen\`eve (Switzerland), ETH Z\"urich (Switzerland), NOVA (Netherlands), ONERA (France) and ASTRON (Netherlands) in collaboration with ESO. SPHERE was funded by ESO, with additional contributions from CNRS (France), MPIA (Germany), INAF (Italy), FINES (Switzerland) and NOVA (Netherlands). SPHERE also received funding from the European Commission Sixth and Seventh Framework Programmes as part of the Optical Infrared Coordination Network for Astronomy (OPTICON) under grant number RII3-Ct-2004-001566 for FP6 (2004-2008), grant number 226604 for FP7 (2009-2012) and grant number 312430 for FP7 (2013-2016).
\end{acknowledgements}

\bibliographystyle{aa}
\bibliography{paper}

\end{document}